\documentclass[twocolumn]{aastex631}

\usepackage[super]{nth}
\usepackage{hyperref}
\accepted{MNRAS March 2023}

\usepackage{color}

\usepackage{url}

\usepackage{xcolor}

\begin{document}

\title{TIC 219006972: A Compact, Coplanar Quadruple Star System Consisting of Two Eclipsing Binaries with an Outer Period of 168 days}

\correspondingauthor{Veselin Kostov}
\email{veselin.b.kostov@nasa.gov}
%
\author[0000-0001-9786-1031]{Veselin~B.~Kostov}
\affiliation{NASA Goddard Space Flight Center, 8800 Greenbelt Road, Greenbelt, MD 20771, USA}
\affiliation{SETI Institute, 189 Bernardo Ave, Suite 200, Mountain View, CA 94043, USA}
\affiliation{GSFC Sellers Exoplanet Environments Collaboration}
\affiliation{Visual Survey Group Collaboration}
%
\author[0000-0002-8806-496X]{Tam\'as Borkovits}
\affiliation{Baja Astronomical Observatory of Szeged University, H-6500 Baja, Szegedi \'ut, Kt. 766, Hungary}
\affiliation{ELKH-SZTE Stellar Astrophysics Research Group, H-6500 Baja, Szegedi \'ut, Kt. 766, Hungary}
\affiliation{Konkoly Observatory, Research Centre for Astronomy and Earth Sciences, H-1121 Budapest, Konkoly Thege Miklós \'ut 15-17, Hungary}
\affiliation{ELTE Gothard Astrophysical Observatory, H-9700 Szombathely, Szent Imre h. u. 112, Hungary}
\affiliation{MTA-ELTE Exoplanet Research Group, H-9700 Szombathely, Szent Imre h. u. 112, Hungary}
%
\author[0000-0003-3182-5569]{Saul A. Rappaport}
\affiliation{Department of Physics, Kavli Institute for Astrophysics and Space Research, M.I.T., Cambridge, MA 02139, USA}
\affiliation{Visual Survey Group Collaboration}
%
\author[0000-0003-0501-2636]{Brian P. Powell}
\affiliation{NASA Goddard Space Flight Center, 8800 Greenbelt Road, Greenbelt, MD 20771, USA}
%
\author[0000-0001-5449-2467]{Andr\'as P\'al}
\affiliation{Konkoly Observatory, Research Centre for Astronomy and Earth Sciences, MTA Centre of Excellence, Konkoly Thege Mikl\'os  \'ut 15-17, H-1121 Budapest, Hungary}
%
%
%
\author[0000-0003-3988-3245]{Thomas L. Jacobs}
\affiliation{Amateur Astronomer, 12812 SE 69th Place, Bellevue, WA 98006}
\affiliation{Visual Survey Group Collaboration}
\author[0000-0002-5665-1879]{Robert Gagliano}
\affiliation{Amateur Astronomer, Glendale, AZ 85308}
\affiliation{Visual Survey Group Collaboration}

\author[0000-0002-2607-138X]{Martti~H.~Kristiansen}
\affil{Brorfelde Observatory, Observator Gyldenkernes Vej 7, DK-4340 T\o{}ll\o{}se, Denmark}
\affiliation{Visual Survey Group Collaboration}
%
\author[0000-0002-8527-2114]{Daryll M. LaCourse}
\affiliation{Amateur Astronomer, 7507 52nd Place NE Marysville, WA 98270}
\affiliation{Visual Survey Group Collaboration}
\author{Maxwell Moe}
\affiliation{Department of Physics \& Astronomy, University of Wyoming, Laramie, WY 82071}
\author{Mark Omohundro}
\affiliation{Citizen Scientist, c/o Zooniverse, Department of Physics, University of Oxford, Denys Wilkinson Building, Keble Road, Oxford, OX13RH, UK}
\affiliation{Visual Survey Group Collaboration}
\author[0000-0002-5034-0949]{Allan R. Schmitt}
\affiliation{Citizen Scientist, 616 W. 53rd. St., Apt. 101, Minneapolis, MN 55419, USA}
%
\author[0000-0002-1637-2189]{Hans M. Schwengeler}
\affiliation{Citizen Scientist, Planet Hunter, Bottmingen, Switzerland}
\affiliation{Visual Survey Group Collaboration}
\author[0000-0002-0654-4442]{Ivan A. Terentev}
\affiliation{Citizen Scientist, Planet Hunter, Petrozavodsk, Russia}
\affiliation{Visual Survey Group Collaboration}
\author[0000-0001-7246-5438]{Andrew Vanderburg}
\affiliation{Department of Astronomy, University of Wisconsin-Madison, Madison, WI 53706, USA}
\affiliation{Visual Survey Group Collaboration}



\begin{abstract}

We present the discovery of a new highly compact quadruple star system, TIC 219006972, consisting of two eclipsing binary stars with orbital periods of 8.3 days and 13.7 days, and an outer orbital period of only 168 days. This period is a full factor of 2 shorter than the quadruple with the shortest outer period reported previously, VW LMi, where the two binary components orbit each other every 355 days. The target was observed by {\em TESS} in Full-Frame Images in sectors 14-16, 21-23, 41, 48 and 49, and produced two sets of primary and secondary eclipses. These show strongly non-linear eclipse timing variations (ETVs) with an amplitude of $\sim$0.1 days, where the ETVs of the primary and secondary eclipses, and of the two binaries are all largely positively correlated. This highlights the strong dynamical interactions between the two binaries and confirms the compact quadruple configuration of TIC 219006972. The two eclipsing binaries are nearly circular whereas the quadruple system has an outer eccentricity of about 0.25. The entire system is nearly edge-on, with a mutual orbital inclination between the two eclipsing binary star systems of about 1 degree.

\end{abstract}

\keywords{stars: binaries (including multiple): close - stars: binaries: eclipsing}


\section{Introduction} \label{sec:intro}

Gravitationally-bound systems of two or more stars are a natural product of star formation and serve as important tracers of the processes and mechanisms driving stellar evolution. When the constituent stars are close enough to each other, the system can experience important (and sometimes quite dramatic) physical and dynamical interactions such as tidal distortions, heating and/or reflection effects, mass transfer, orbital oscillations and/or perturbations, common-envelope events, collisions and even supernovae explosions \citep[e.g.][]{Lidov1962,1962AJ.....67..591K,2013MNRAS.435..943P,2009ApJ...697.1048P,2021MNRAS.502.4479H,2014ApJ...793..137N,2018MNRAS.476.4234F,2019MNRAS.486.4781F,2019MNRAS.483.4060L,2022Galax..10....9B}.

Multiple stellar systems cover a vast range of occurrence rates, orbital configurations, physical parameters, number of components, ages, etc. Overall, about one in every ten Sun-like stars resides in a triple system and a few percent are members of quadruple or higher-order systems \citep[e.g][]{Raghavan2010,DeRosa2014,Toonen2016,Moe2017,Tokovinin2017}. 

Compact multiple star systems are of particular interest for studying dynamical relationships. For example, the current record holders for the confirmed shortest outer periods in triple and quadruple systems are: (i) the 2+1 triple system $\lambda$ Tau with $P_{out} = 33.03$ days \citep{1916PAllO...3...23S,1956ApJ...124..507E}; (ii) the 2+1+1 quadruple system TIC 114936199 with $P_{middle}\sim 51$ days \citep{2022ApJ...938..133P}; and (iii) the 2+2 quadruple system VW LMi where the two binary components orbit each other every 355 days \citep{Pribulla2008,Pribulla2020}. The outer period of TIC 219006972 is about half as long as that of VW LMi, making it the most compact 2+2 quadruple system reported to date\footnote{We note that the quadruple system candidate BU CMi (TIC 271204362), announced by Jayaraman et al. (in prep.) at the {\em TESS} Science Conference II in 2021, has an outer orbital period of about 120 days. This is even shorter than the outer period of TIC 219006972, which will then make it only the second quadruple with an outer period of $\lesssim 1/2$ year.}.

Stellar multiples that contain one or more eclipsing binary stars, and have an outer period short enough for detectable interactions between the component sub-systems provide excellent ``hunting grounds'' for discovery and detailed characterization of triple and higher-order systems. Large-scale photometric surveys are well-positioned to observe such compact multi-stellar systems. For example, NASA's Transiting Exoplanet Survey Satellite ({\em TESS}) mission \citep{Ricker14} is expected to observe hundreds of thousands of eclipsing binary stars \citep{2015ApJ...809...77S} and provide high-precision, nearly-continuous month-long observations spanning multiple epochs. Indeed, {\em TESS} already demonstrated its potential for stellar multiples by enabling the discovery of hundreds of triple and quadruple systems \citep[e.g.][]{2022MNRAS.513.4341R,2022MNRAS.510.1352B,2022A&A...664A..96Z,2022ApJS..259...66K}, and even the first two fully eclipsing sextuple systems \citep{2021AJ....161..162P,2023MNRAS.tmp..346Z}.

Here we present the latest addition to the still-small family of compact quadruple systems, TIC 219006972, composed of two eclipsing binaries. At the time of writing, TIC 219006972 has the shortest outer orbital period reported, and is the second closest to co-planarity (after TIC 454140642). This paper is organized as follows. Section 2 describes the detection and preliminary analysis of the system. In Section 3, we outline the comprehensive photometric-dynamical solution of the system, followed by a discussion of its properties in Section 4. Section 5 summarizes our results. 

\section{Detection}

\begin{figure}
    \centering
    \includegraphics[width=0.99\columnwidth]{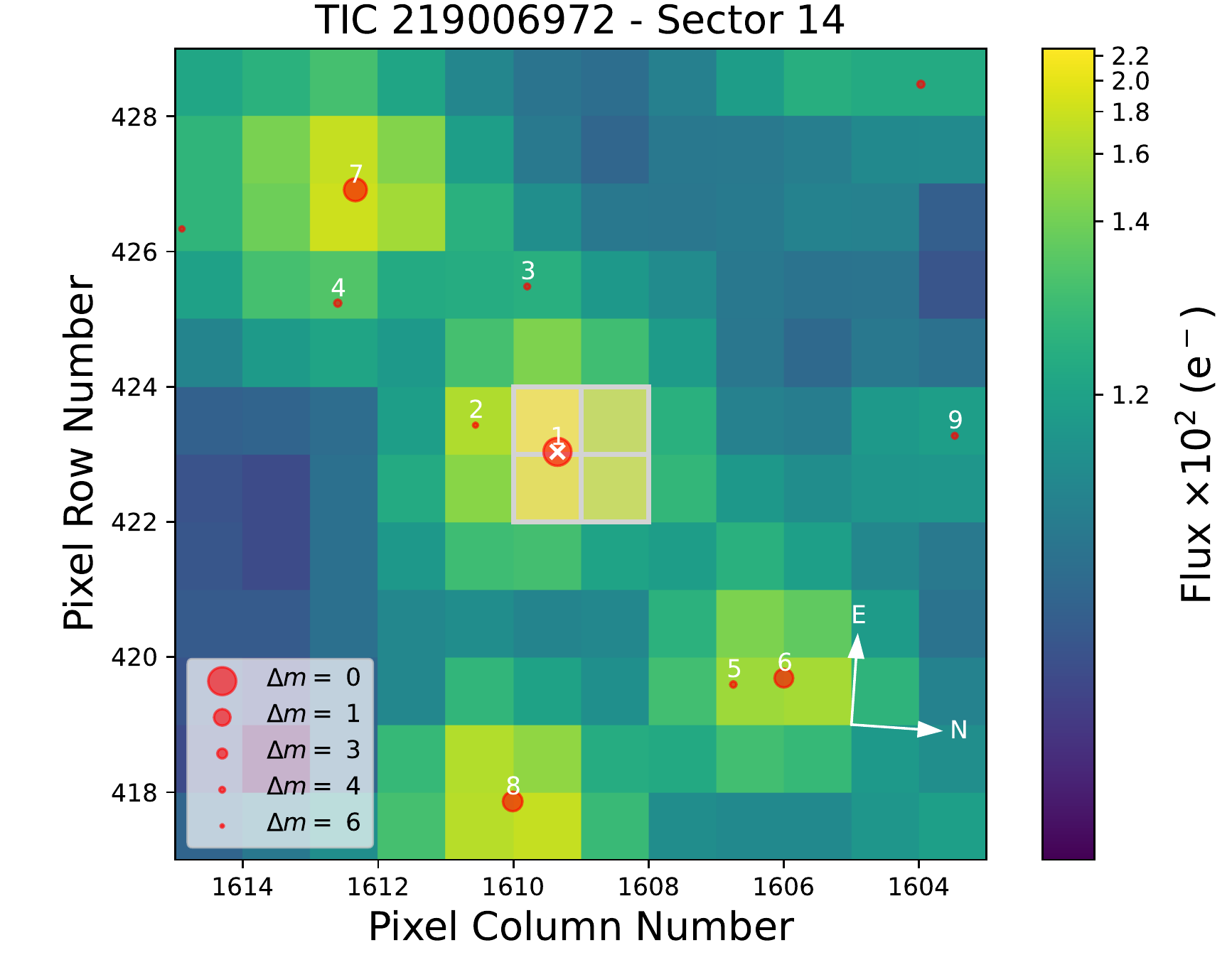}
    \caption{$12 \times 12$ {\em TESS} pixels image of TIC 219006972 (marked with a cross-hair) for Sector 14 showing all nearby stars brighter than a {\em TESS} magnitude difference ${\Delta m = 6\ \rm mag}$. Image created with \texttt{tpfplotter} \citep{2020A&A...635A.128A}.}
    \label{fig:Skyview}
\end{figure}

TIC 219006972 is a relatively faint (V = 14.312 mag) star in the Northern hemisphere (RA = 14:44:48.71, Dec = +66:22:43.17) with a Gaia DR3 parallax of 6.65 mas/yr, effective temperature of 5651 K, zero $astrometric\_excess\_noise$, and RUWE = 1.01 \citep{Gaia2021}. The target is relatively isolated -- the closest field star is TIC 1102519834, 17.8 arcsec away and $\Delta$G = 6.2 mag fainter (see Fig. \ref{fig:Skyview}), with another 6 sources within 1 arcmin radial search, all $\Delta$G = 4.5 mag and fainter. The parameters of the system are listed in Table 1. 

\begin{deluxetable}{l r r r }[!ht]
\tabletypesize{\scriptsize}
\tablecaption{Stellar parameters for TIC 219006972\label{tab:EBparameters}}
\tablewidth{0pt}
\tablehead{
\colhead{Parameter} & \colhead{Value} & \colhead{Error} &\colhead{Source}
}
\startdata
\multicolumn{4}{l}{\bf Identifying Information} \\
\hline
TIC ID & 219006972 & & 1 \\
Gaia ID & 1669481894221366912 & & 2 \\
$\alpha$  (J2000, hh:mm:ss) & 14:44:48.71 &  & 1 \\
$\delta$  (J2000, dd:mm:ss) & +66:22:43.17 &  & 1 \\
$\mu_{\alpha}$ (mas~yr$^{-1}$) & $6.6476$ & 0.0174 & 2 \\
$\mu_{\delta}$ (mas~yr$^{-1}$) & $-3.4518$ & 0.01511 & 2 \\
$\varpi$ (mas) & 0.7313 & 0.0133 & 2 \\
Distance (pc) & 1367 & 32 & 2\\
RUWE & 1.0121 & & 2 \\
\\
\multicolumn{4}{l}{\bf Photometric Properties} \\
\hline
$T$ (mag) & 13.7021 & 0.0063 & 1 \\
$B$ (mag) & 14.9 & 0.106 & 1\\
$V$ (mag) & 14.312 & 0.172 & 1 \\
$Gaia$ (mag) & 14.1153 &  0.0028 & 2 \\
$J$ (mag) & 13.073 & 0.025 & 3 \\
$H$ (mag) & 12.75 & 0.025 & 3 \\
$K$ (mag) & 12.676 & 0.025 & 3 \\
$W1$ (mag) & 12.686 & 0.022 & 4 \\
$W2$ (mag) & 12.705 & 0.023 & 4 \\
$W3$ (mag) & 12.582 & 0.292 & 4 \\
$W4$ (mag) & 9.56 & & 4 \\
\\
\multicolumn{4}{l}{\bf Stellar Properties} \\
\hline
Teff, Aa (K) & 6162 & 90 & This work\\
Teff, Ab (K) & 5942 & 80 & This work\\
Teff, Ba (K) & 5676 & 110 & This work\\
Teff, Bb (K) & 3698 & 47 & This work\\
$\lbrack $Fe/H$ \rbrack$  & -0.275 & 0.076 & This work \\
Age (Gyr) & 6 & 1.5 & This work\\
\enddata
Sources: (1) TIC-8 \citep{TIC}, (2) Gaia DR3 \citep{Gaia2021}, (3) 2MASS All-Sky Catalog of Point Sources \citep{2MASS}, (4) AllWISE catalog \citep{WISE}
\tablenotetext{}{}
\end{deluxetable}

\begin{figure*}
    \centering
    \includegraphics[width=0.92\linewidth]{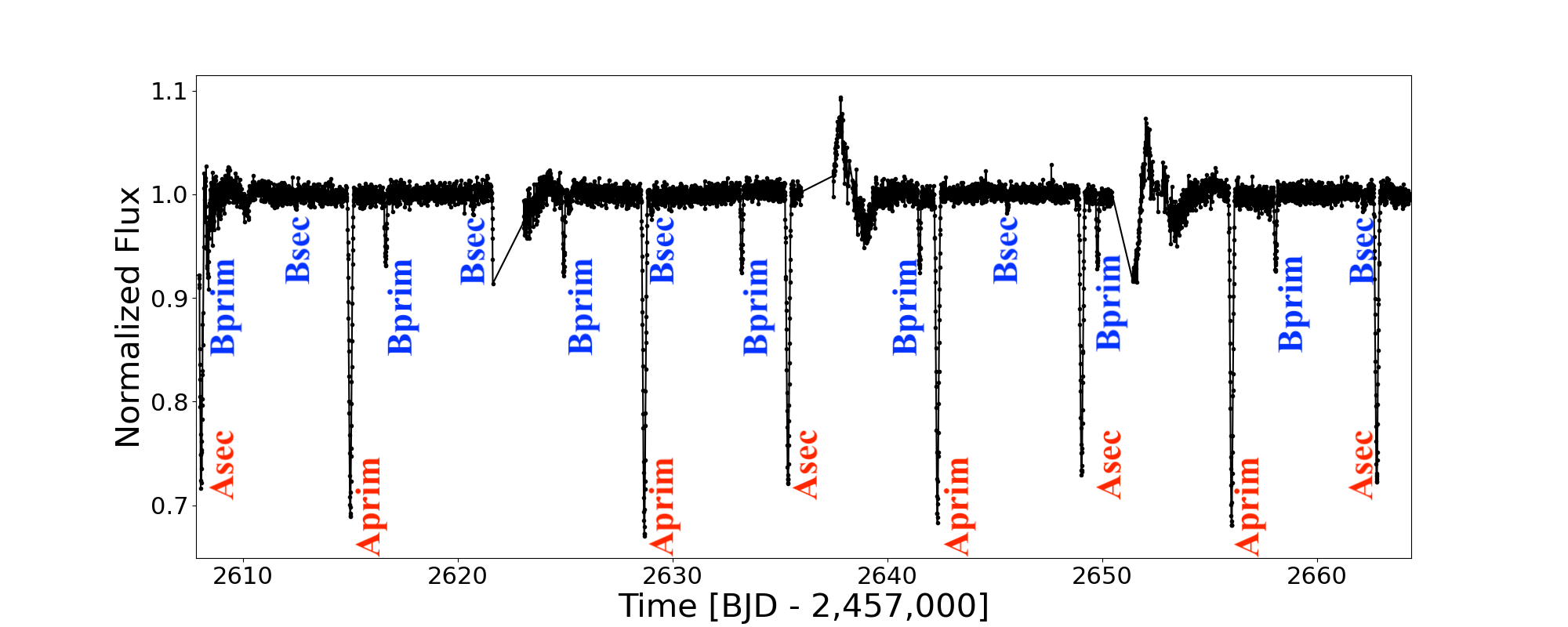}
    \caption{{\em TESS} Full-Frame-Image (FFI) \texttt{eleanor} Corrected Flux photometry from Sectors 48 and 49, highlighting select primary and secondary eclipses for both binaries. }
    \label{fig:TESS_lc}
\end{figure*}

{\em TESS} observed TIC 219006972 in sectors 14, 15, 16, 21, 22, 23, 41, 48, and 49. The long-cadence {\em TESS} photometry for Sectors 48 and 49 is shown in Figure \ref{fig:TESS_lc}, highlighting the two sets of eclipses from binary A (period = 13.7 days) and binary B (period = 8.3 days). We note that there are no clear signatures of stellar rotation in the long-cadence lightcurves. 

TIC 219006972 was initially identified as a target of interest through visual survey of a data set of lightcurves with a high probability of containing eclipses generated from the {\em TESS} sector 48 data release. Briefly, upon the release of the Full Frame Images (FFI) from each {\em TESS} sector, we download the FFI data and construct all lightcurves brighter than \nth{15} magnitude using the {\em Discover} supercomputer at the NASA Center for Climate Simulation.\footnote{\url{https://www.nccs.nasa.gov/systems/discover}} We then process the lightcurves through a neural network trained to identify eclipses, further described in \cite{2022RNAAS...6..111P}. We selected the lightcurves identified by the neural network as likely containing eclipses, and distributed them to the Visual Survey Group \citep[VSG,][]{2022PASP..134g4401K} for visual review. 

\begin{figure*}
    \centering
    \includegraphics[width=0.95\linewidth]{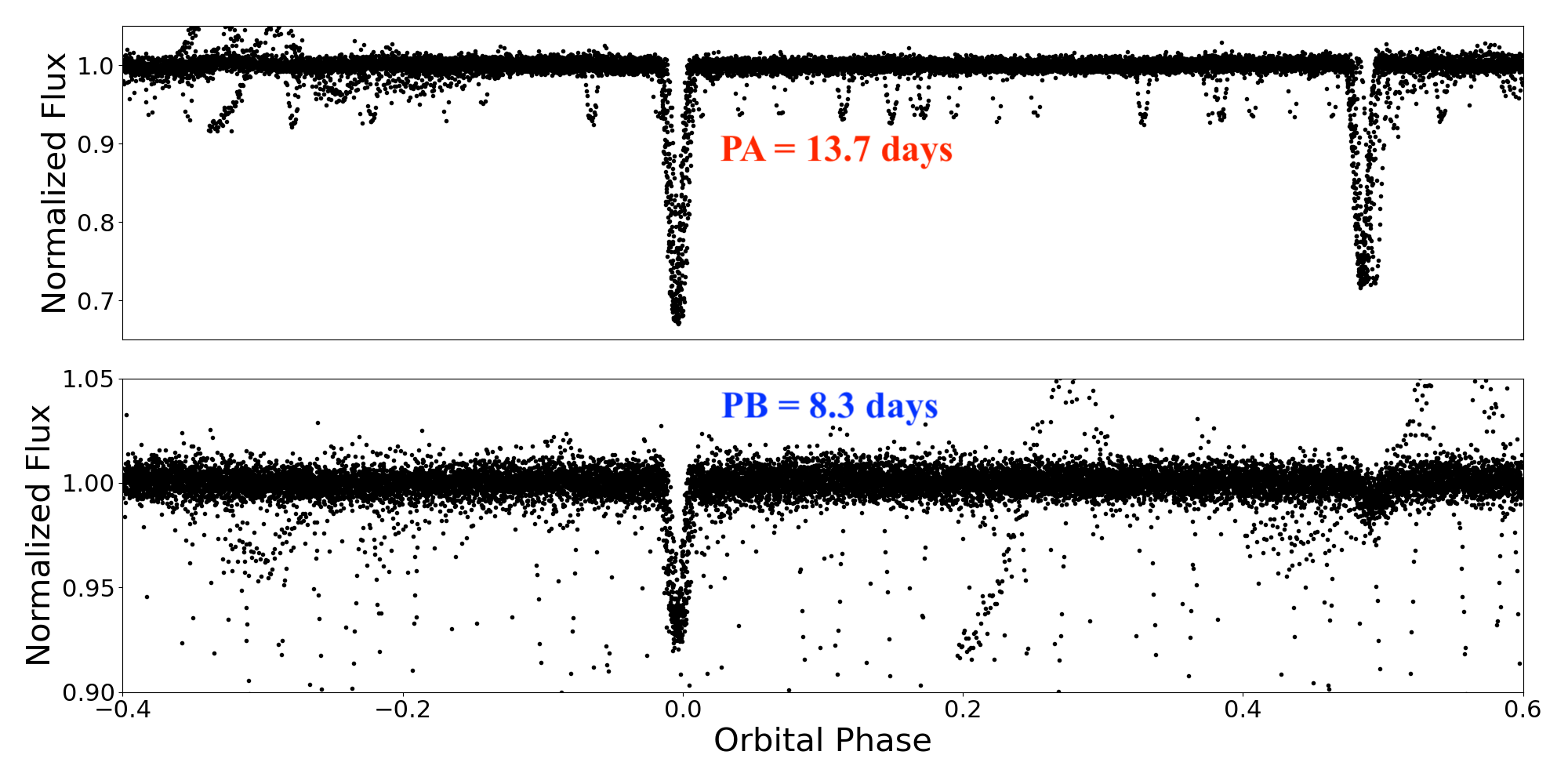}
    \caption{Phase-folded {\em TESS} \texttt{eleanor} data on the two periods of PA = 13.65 days (upper panel) and PB = 8.3 days (lower panel), highlighting the deviations from linear ephemeris for both binaries.}
    \label{fig:folded_lc_TESS}
\end{figure*}

\begin{figure}
    \centering
    \includegraphics[width=0.99\columnwidth]{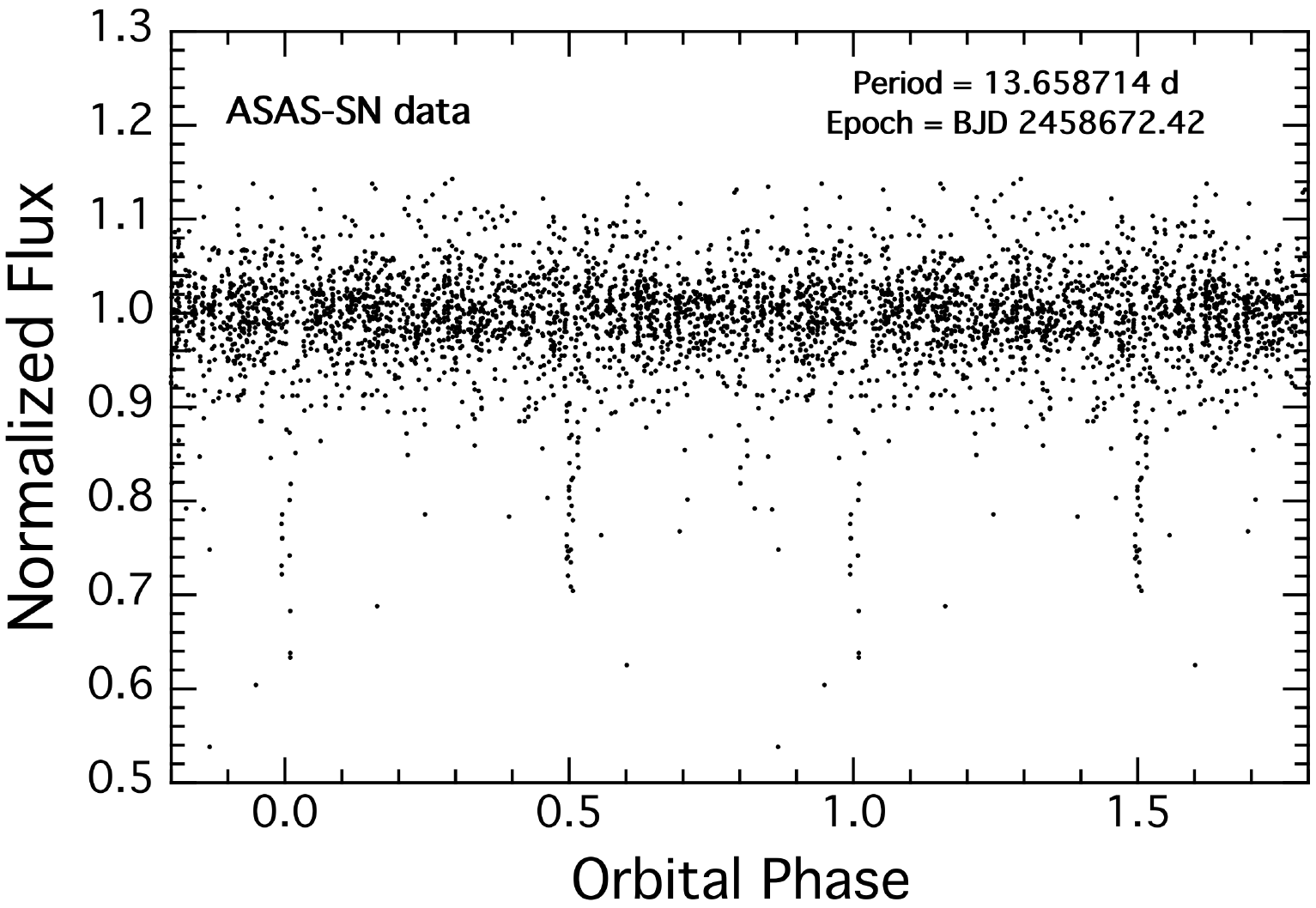}
    \includegraphics[width=0.99\columnwidth]{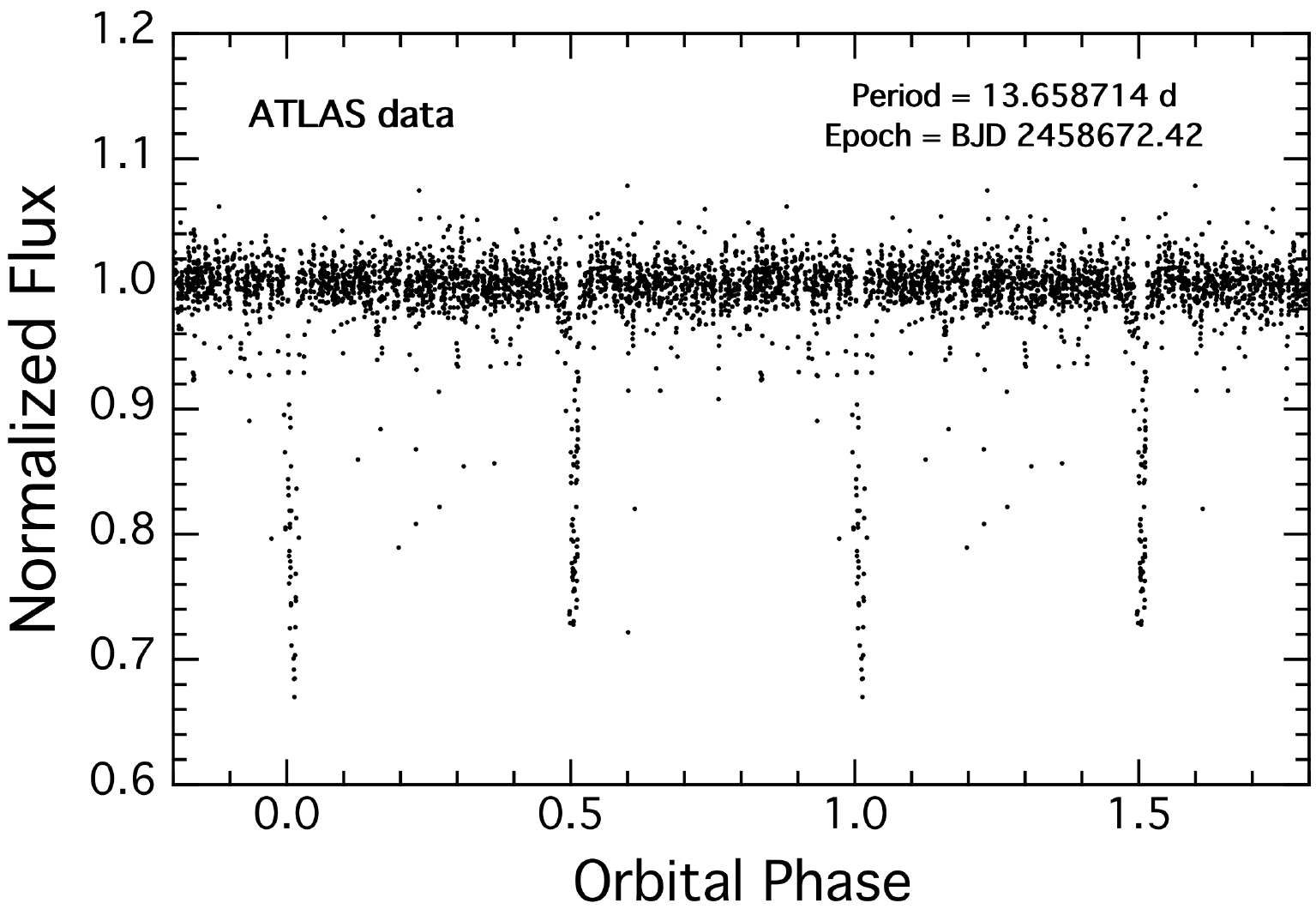}
    \caption{Phase-folded ASAS-SN (upper panel) and ATLAS (lower panel) data for TIC 219006972. The PA = 13.65-day eclipses are clearly seen in both the ASAS-SN and the ATLAS photometry, and demonstrate that there are no drastic changes in the PA binary between {\em TESS} and ASAS-SN/ATLAS.}
    \label{fig:folded_lc_archive}
\end{figure}

Using the LcTools software system \citep{2019arXiv191008034S,2021arXiv210310285S}, members of the VSG identified TIC 219006972 as a likely multiple star system in December 2022 with periods of PA = 13.7 days and PB = 8.3 days (upper panels, Fig. \ref{fig:folded_lc_TESS}), and the target was accordingly promoted for additional analysis. First, an archival search showed that TIC 219006972 was observed as part of the ASAS-SN \citep{2017PASP..129j4502K} and ATLAS \citep{2018AJ....156..241H} projects. The PA eclipses are detected in both datasets (see Fig. \ref{fig:folded_lc_archive}), demonstrating that there are no drastic long-term changes in the PA binary.

\begin{figure}
    \centering
    \includegraphics[width=0.45\linewidth]{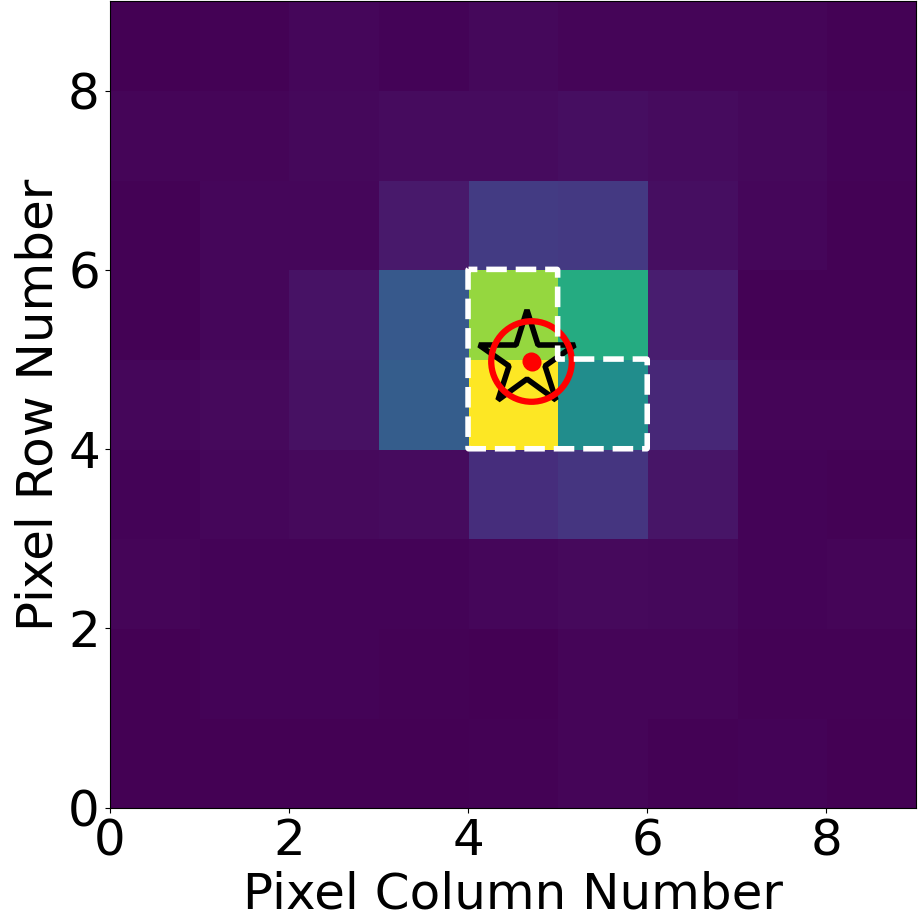}
    \includegraphics[width=0.45\linewidth]{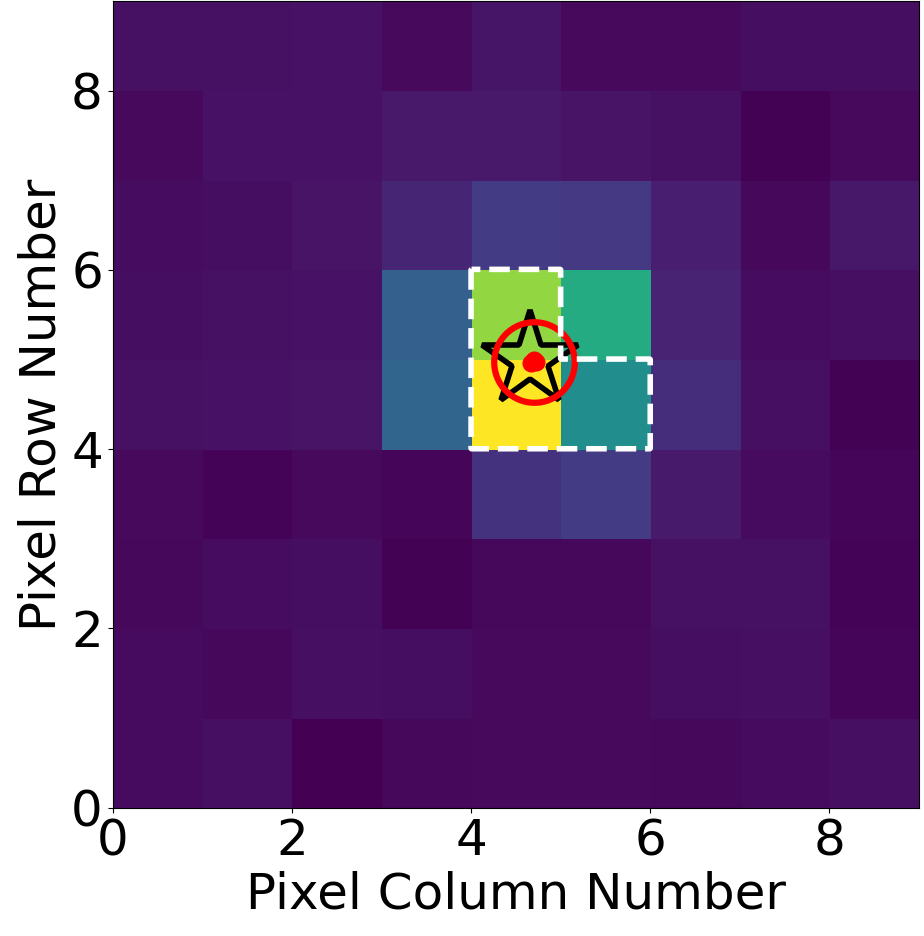}
    \caption{Difference images showing center-of-light measurements for PA (left) and PB (right) during the detected eclipses in Sector 41. The star symbols represent the catalog position of the target, the small red dots represent the individual center-of-light measurements for each detected eclipses, the large red circle represents their average, and the dashed white contour represents the pixel aperture used by Eleanor to extract the lightcurve. The measurements confirm that TIC 219006972 is the source of both PA and PB.}
    \label{fig:photocenters}
\end{figure}

Second, detailed investigation of the motion of the center-of-light during the two sets of eclipses\footnote{Details of the procedure are described in \cite{2021ApJ...917...93K}.} confirmed that they originate from the target (see Figure \ref{fig:photocenters}), in line with its isolated nature, further strengthening the quadruple hypothesis. 

\begin{figure*}
    \centering
    \includegraphics[width=0.98\linewidth]{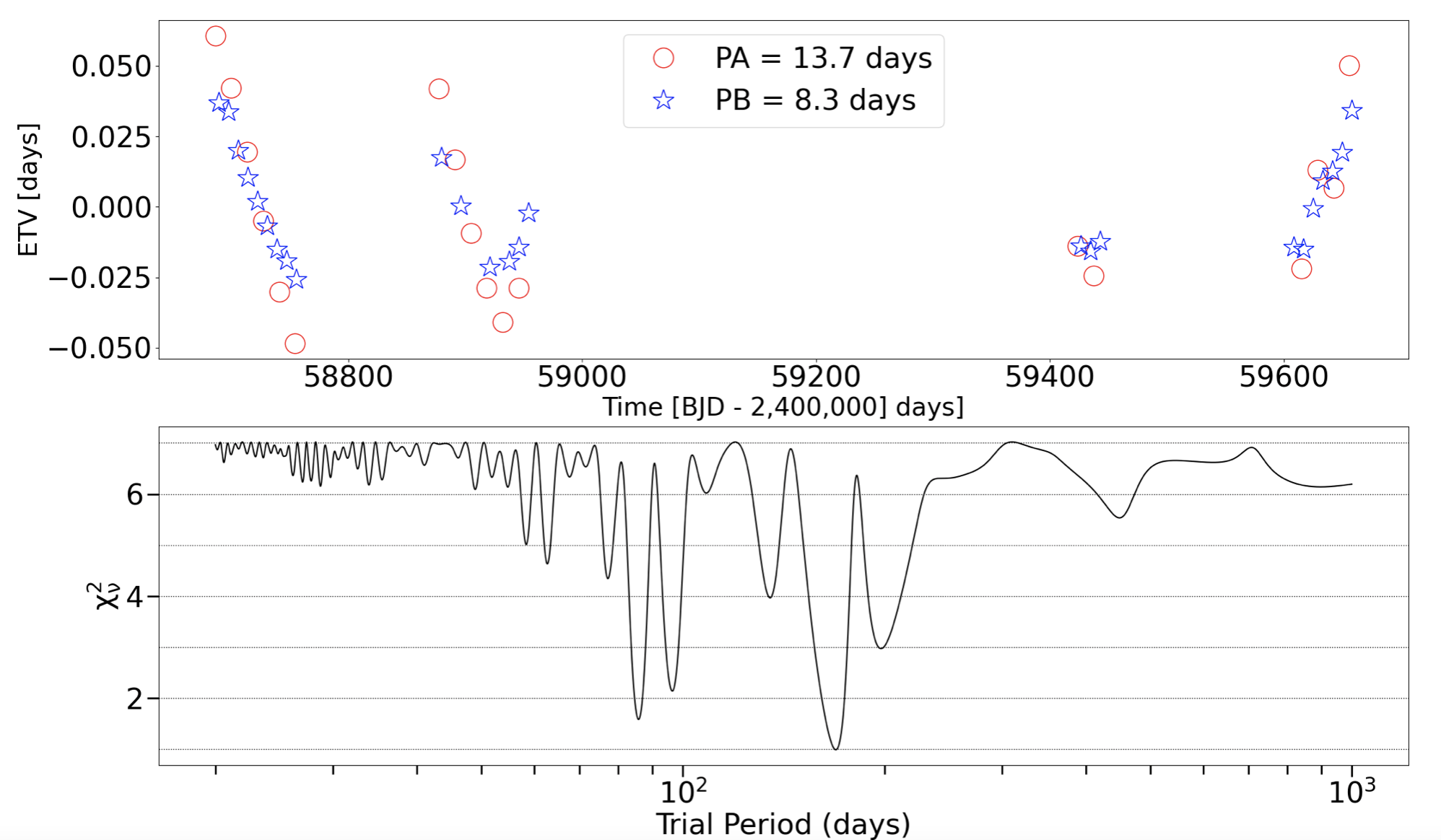}
    \caption{First panel: Preliminary primary ETVs for the PA (red circles) and PB (blue stars) binaries based on the {\texttt{eleanor}} lightcurve showing clear, correlated deviations from a linear ephemeris. Second panel: $\chi^2$ fits to the measured ETVs using simple sinusoids with a large set of oversampled periods. This indicates an initial estimate for the outer period of 169 days.}
    \label{fig:vetting}
\end{figure*}

Next, a preliminary analysis of the {\em TESS} lightcurve revealed prominent non-linear and correlated Eclipse Timing Variations (ETVs) with a large total amplitude of about 0.1 days (see the upper panel of Fig.~\ref{fig:vetting}). In order to check for any obvious periodicity in the ETV curves we carried out $\chi^2$ fits using simple sinusoids with a large set of oversampled periods. While most ETV curves, due to either Light Travel Time Effects (LTTE) or dynamical delays (see Sect.~\ref{sec:photodynamics}), are distinctly non-sinusoidal, the dominant Fourier component is still either at the orbital period or its half period---so this is a useful first check for any periodicity.  The bottom panel in Fig.~\ref{fig:vetting} shows the resultant $\chi^2$ value vs.~the trial orbital period when fitting these sinusoids to the ETV curve above.  The deepest $\chi^2$ minimum (best fit) comes at $P_{\rm out} = 169.3 \pm 2$ days.  The next deepest minimum, at 169/2 days, represents a higher harmonic of $P_{\rm out}$, and can be ruled out as the most significant minimum at the 5-$\sigma$ level.  The third deepest minimum occurs at a period of 96 days and is excluded as the most significant minimum at the 7-$\sigma$ level. Thus, we tentatively conclude that the outer period of this quadruple is 169 days, which would comprise the shortest outer orbital period for a published quadruple star system. For a much more comprehensive analysis of the system and derivation of its stellar and orbital parameters we employed a detailed photometric-dynamical model as described in the next section. 

\section{Photodynamical analysis of the system}
\label{sec:photodynamics}

We carried out a simultaneous, joint analysis of the \textit{TESS} lightcurve, the ETV curves derived from the mid-times of the \textit{TESS}-observed eclipses, and the available multi-passband SED data of the system with the {\sc Lightcurvefactory} software package \citep{Borkovits2019,Borkovits2020}. During the analysis we followed the same steps which were described in detail in our previous analysis of TIC~454140642 \citep{2021ApJ...917...93K}, another \textit{TESS}-discovered compact quadruple system. The only fundamental difference between the two studies is, that in contrast to this latter system, no Radial Velocity (RV) observations are available for TIC~219006972 and, hence, we were not able to fit RV curve(s) simultaneous to the other parts of the analysis. 
For further details on the photodynamical analysis we refer the reader to Sect.~5.1 of \citet{2021ApJ...917...93K}. Here, we describe only the data preparation steps which are specific for this system.

As mentioned previously, TIC~219006972 was observed in six sectors during Year 2, and three sectors during Year 4 observations. For the photodynamical analysis we extracted 30-min and 10-min cadenced lightcurves from the FFI files with the {\sc Fitsh} package \citep{2012MNRAS.421.1825P}. We carried out low-order polynomial detrending on the out-of-eclipse sections to eliminate the most trivial, instrumental systematics. Finally, to reduce the computational time, we removed the out-of-eclipse lightcurve sections, keeping only narrow, $\pm0\fp05$-phase-domain regions centered on each eclipse.

\begin{deluxetable*}{lrllrllrllrl }[!ht]
\tabletypesize{\scriptsize}
\tablecaption{Mid-eclipse Times for TIC 219006972A\label{Tab:TIC219006972A_ToM}}
\tablewidth{0pt}
\tablehead{
\colhead{BJD} & \colhead{Cycle} & \colhead{std. dev.} & \colhead{BJD} & \colhead{Cycle} & \colhead{std. dev.} & \colhead{BJD} & \colhead{Cycle} & \colhead{std. dev.} & \colhead{BJD} & \colhead{Cycle} & \colhead{std. dev.}}

\startdata
$-2\,400\,000$ & no. &   \multicolumn{1}{c}{$(d)$} & $-2\,400\,000$ & no. &   \multicolumn{1}{c}{$(d)$} & $-2\,400\,000$ & no. &   \multicolumn{1}{c}{$(d)$} & $-2\,400\,000$ & no. &   \multicolumn{1}{c}{$(d)$} \\ 
\hline
58686.282316 &    0.0 & 0.000852 & 58747.534495 &    4.5 & 0.000765 & 58925.077543 &   17.5 & 0.000700 & 59608.037828 &   67.5 & 0.000547  \\ 
58692.978462 &    0.5 & 0.000756 & 58754.466837 &    5.0 & 0.000829 & 58932.035148 &   18.0 & 0.000672 & 59614.985457 &   68.0 & 0.000332  \\ 
58699.923404 &    1.0 & 0.000823 & 58761.184269 &    5.5 & 0.000560 & 58938.733397 &   18.5 & 0.000521 & 59628.679520 &   69.0 & 0.000390  \\ 
58706.618686 &    1.5 & 0.000500 & 58870.515620 &   13.5 & 0.001813 & 58945.707518 &   19.0 & 0.000737 & 59635.372093 &   69.5 & 0.000229  \\ 
58713.559120 &    2.0 & 0.000512 & 58877.484039 &   14.0 & 0.000571 & 58952.406999 &   19.5 & 0.000854 & 59642.331523 &   70.0 & 0.000275  \\ 
58720.255316 &    2.5 & 0.000699 & 58891.118605 &   15.0 & 0.000641 & 59423.772646 &   54.0 & 0.000358 & 59649.047233 &   70.5 & 0.000263  \\ 
58727.193553 &    3.0 & 0.000734 & 58904.752666 &   16.0 & 0.000414 & 59430.453162 &   54.5 & 0.000328 & 59656.033032 &   71.0 & 0.000287  \\ 
58733.894683 &    3.5 & 0.000504 & 58911.431342 &   16.5 & 0.000675 & 59437.421320 &   55.0 & 0.000283 & 59662.792741 &   71.5 & 0.000267  \\ 
58740.828295 &    4.0 & 0.000701 & 58918.389326 &   17.0 & 0.000616 & 59444.119502 &   55.5 & 0.000256  \\ 
\enddata
\tablenotetext{}{{\em Notes.} Integer and half-integer cycle numbers refer to primary and secondary minima, respectively}
\end{deluxetable*}

\begin{deluxetable*}{lrllrllrllrl }[!ht]
\tabletypesize{\scriptsize}
\tablecaption{Mid-Eclipse Times for TIC 219006972B\label{Tab:TIC219006972B_ToM}}
\tablewidth{0pt}
\tablehead{
\colhead{BJD} & \colhead{Cycle} & \colhead{std. dev.} & \colhead{BJD} & \colhead{Cycle} & \colhead{std. dev.} & \colhead{BJD} & \colhead{Cycle} & \colhead{std. dev.} & \colhead{BJD} & \colhead{Cycle} & \colhead{std. dev.}}

\startdata
$-2\,400\,000$ & no. &   \multicolumn{1}{c}{$(d)$} & $-2\,400\,000$ & no. &   \multicolumn{1}{c}{$(d)$} & $-2\,400\,000$ & no. &   \multicolumn{1}{c}{$(d)$} & $-2\,400\,000$ & no. &   \multicolumn{1}{c}{$(d)$} \\ 
\hline
58685.236288 &   -0.5 & 0.002285 & 58738.971483 &    6.0 & 0.001241 & 58917.007916 &   27.5 & 0.002995 & 59608.349704 &  111.0 & 0.000646  \\ 
58689.344931 &    0.0 & 0.000406 & 58743.145273 &    6.5 & 0.005062 & 58921.123356 &   28.0 & 0.000697 & 59612.434059 &  111.5 & 0.004515  \\ 
58693.513113 &    0.5 & 0.002803 & 58747.245300 &    7.0 & 0.000273 & 58925.309371 &   28.5 & 0.154352 & 59616.629119 &  112.0 & 0.000224  \\ 
58697.620699 &    1.0 & 0.000559 & 58755.521954 &    8.0 & 0.000379 & 58929.397837 &   29.0 & 0.000447 & 59620.719734 &  112.5 & 0.001450  \\ 
58705.889871 &    2.0 & 0.000448 & 58871.491234 &   22.0 & 0.000353 & 58937.682106 &   30.0 & 0.000301 & 59624.922346 &  113.0 & 0.000276  \\ 
58710.059326 &    2.5 & 0.002269 & 58875.649635 &   22.5 & 0.001726 & 58945.969802 &   31.0 & 0.000576 & 59628.996033 &  113.5 & 0.001859  \\ 
58714.159819 &    3.0 & 0.000940 & 58879.761833 &   23.0 & 0.000331 & 58954.259163 &   32.0 & 0.000233 & 59633.211735 &  114.0 & 0.000251  \\ 
58718.323564 &    3.5 & 0.003754 & 58888.031673 &   24.0 & 0.000305 & 59422.031984 &   88.5 & 0.001088 & 59641.495204 &  115.0 & 0.000229  \\ 
58722.424307 &    4.0 & 0.010602 & 58892.185449 &   24.5 & 0.008371 & 59426.196701 &   89.0 & 0.000479 & 59645.588546 &  115.5 & 0.004204  \\ 
58726.600576 &    4.5 & 0.001527 & 58896.303106 &   25.0 & 0.000371 & 59434.474604 &   90.0 & 0.000371 & 59649.777672 &  116.0 & 0.000209  \\ 
58730.699461 &    5.0 & 0.000258 & 58900.453999 &   25.5 & 0.003658 & 59438.564007 &   90.5 & 0.002448 & 59658.080745 &  117.0 & 0.000253  \\ 
58734.876726 &    5.5 & 0.008773 & 58908.734092 &   26.5 & 0.006369 & 59442.759324 &   91.0 & 0.000355 & 59662.180871 &  117.5 & 0.001639  \\ 
\enddata
\tablenotetext{}{{\em Notes.} Integer and half-integer cycle numbers refer to primary and secondary minima, respectively}
\end{deluxetable*}

Additionally, we used the four ETV curves (primary and secondary ETV data for both binaries; see Tables~\ref{Tab:TIC219006972A_ToM} and \ref{Tab:TIC219006972B_ToM}). Furthermore, the observed passband magnitudes tabulated in Table~\ref{tab:EBparameters} were also used for the SED analysis. Similar to our previous works, for the SED analysis we used a minimum uncertainty of $0.03$ mag for most of the observed passband magnitudes, in order to avoid the outsized contribution of the extremely precise Gaia magnitudes and also to counterbalance the uncertainties inherent in our interpolation method during the calculations of theoretical passband magnitudes that are part of the fitting process. The two exceptions are the WISE $W4$ and GALEX $FUV$ magnitudes, for which the uncertainties were inflated to 0.3 mag.

\begin{figure*}
    \centering
    \includegraphics[width=0.65\linewidth]{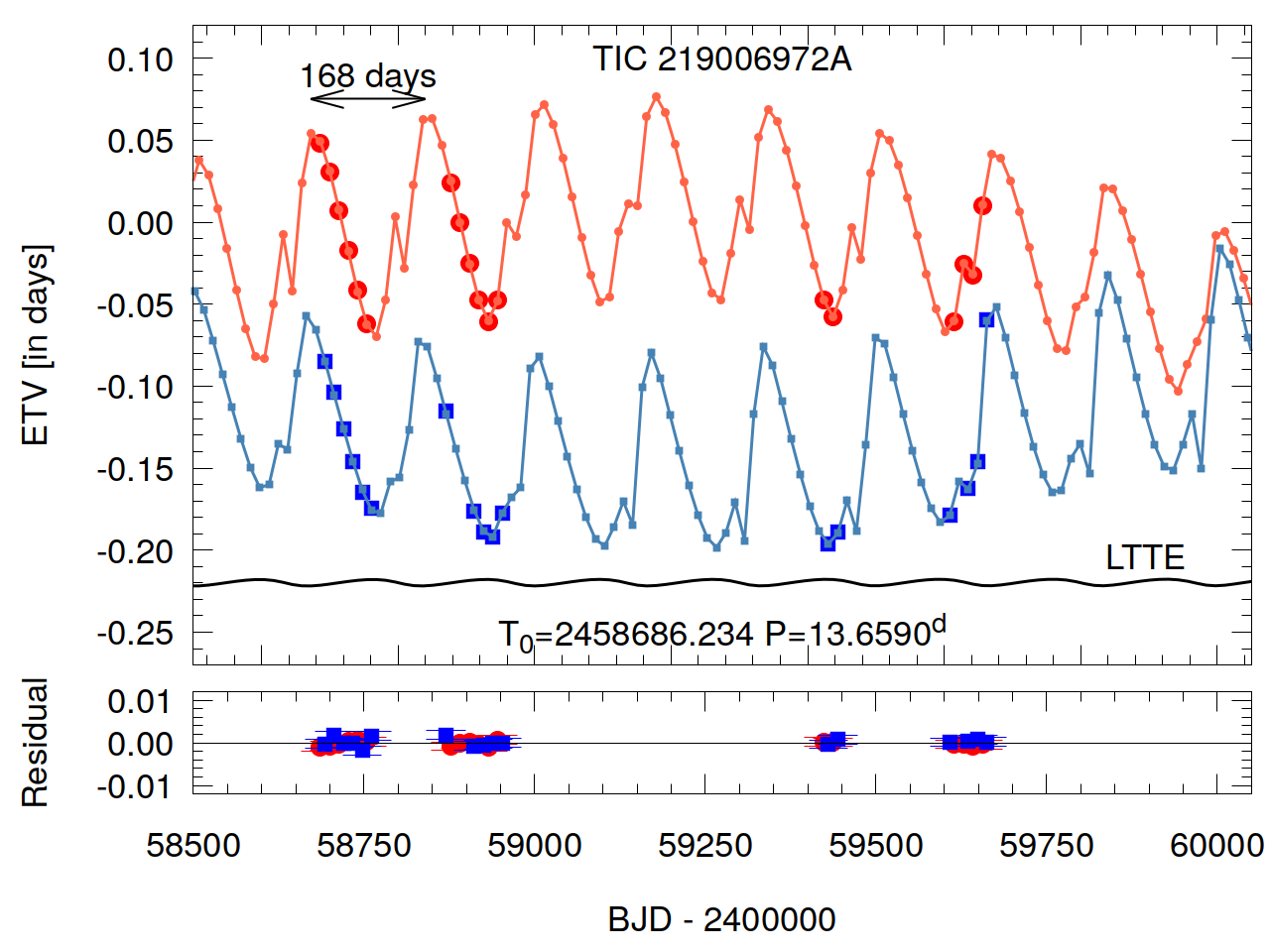}
    \includegraphics[width=0.65\linewidth]{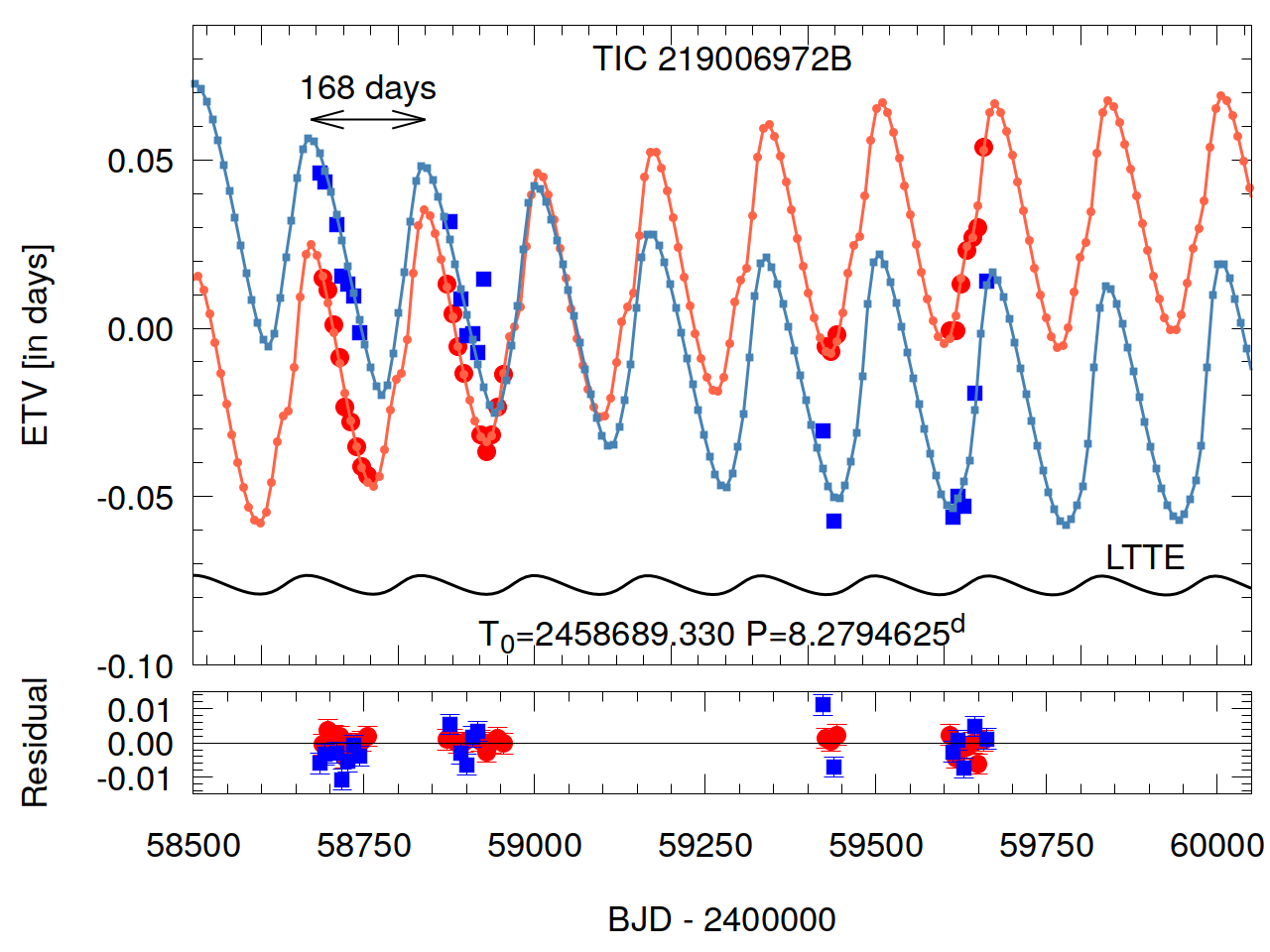}
    \caption{Photodynamical solution for eclipse timing variations of TIC 219006972 showing the best-fit model and corresponding residuals (top panel for binary A, bottom panel for binary B). The red and blue symbols denote the measured primary and secondary ETVs, respectively, while the thin red and blue dots, connected with straight lines, represent the corresponding model. The thin black curves represent the LTTE contributions to the ETVs. As clearly seen from the figure, the dynamical perturbations completely dominate the measured eclipse times.}
    \label{fig:ETVlcfit}
\end{figure*}

\begin{figure}
    \centering
    \includegraphics[width=0.99\columnwidth]{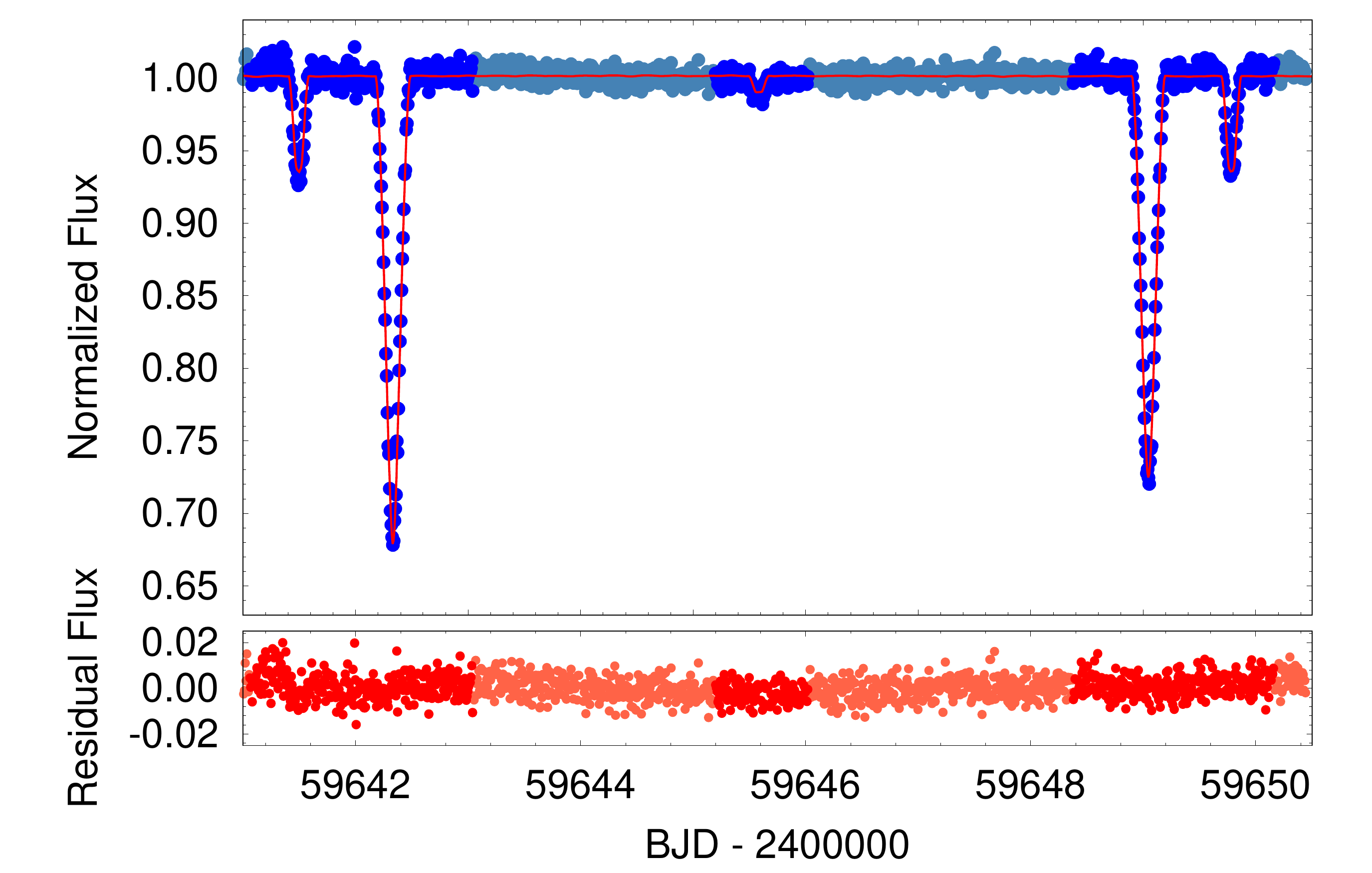}
    \includegraphics[width=0.99\columnwidth]{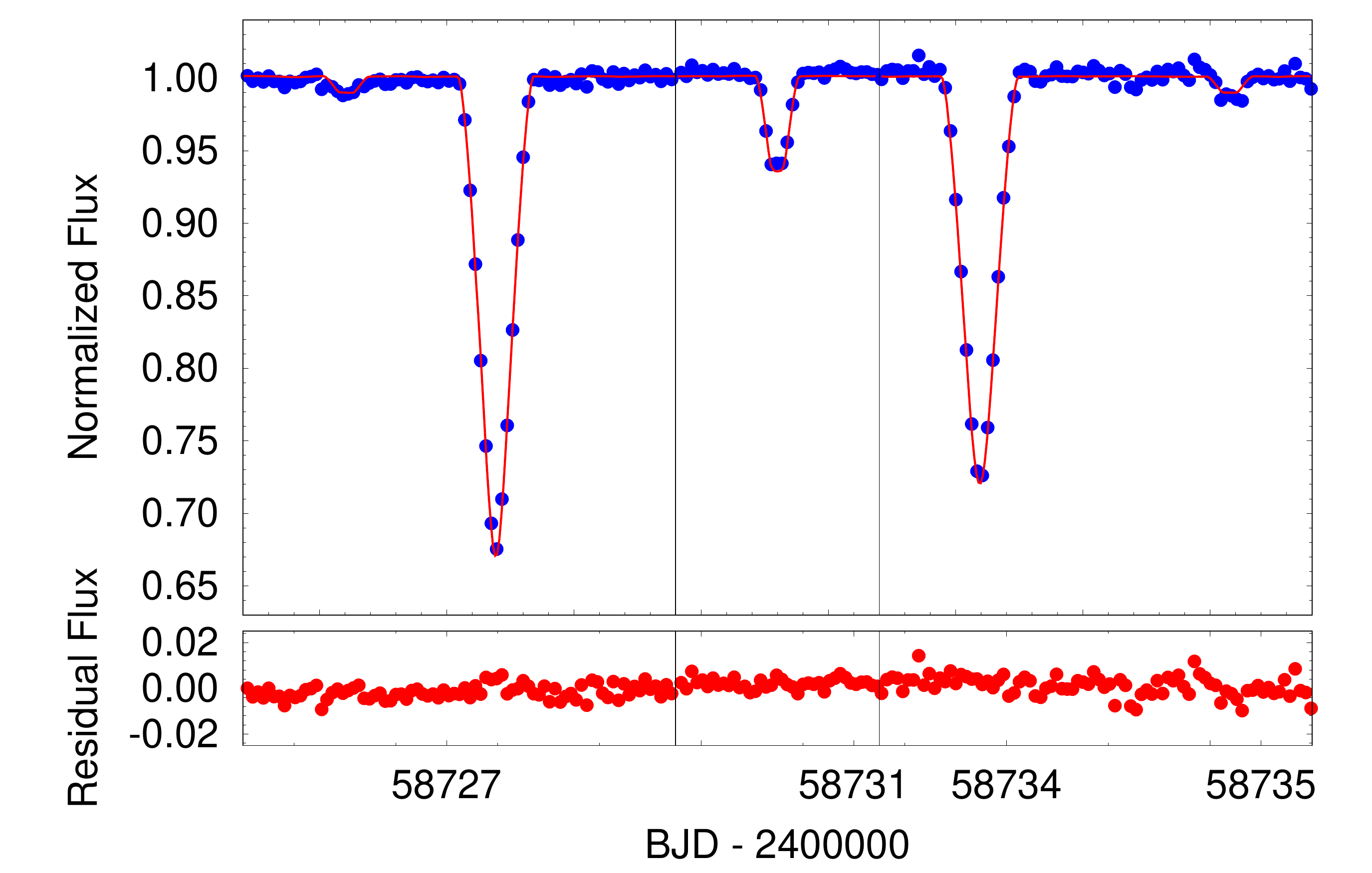}
    \caption{Example sections of the \textit{TESS} lightcurve (blue points) along with the best-fit model (red) and the corresponding residuals. The upper panel represents a continuous, 9-day-long section of the \textit{TESS} lightcurve. Here the pale blue points (and the corresponding orange residual points) show those out-of-eclipse data (and their residuals) which were not used for the photodynamical analysis. The lower panel represent three different sections of the lightcurve for a beter view of the four different kind of eclipses.}
    \label{fig:ETVlcfit2}
\end{figure}

\begin{figure}
    \centering
    \includegraphics[width=1.02\linewidth]{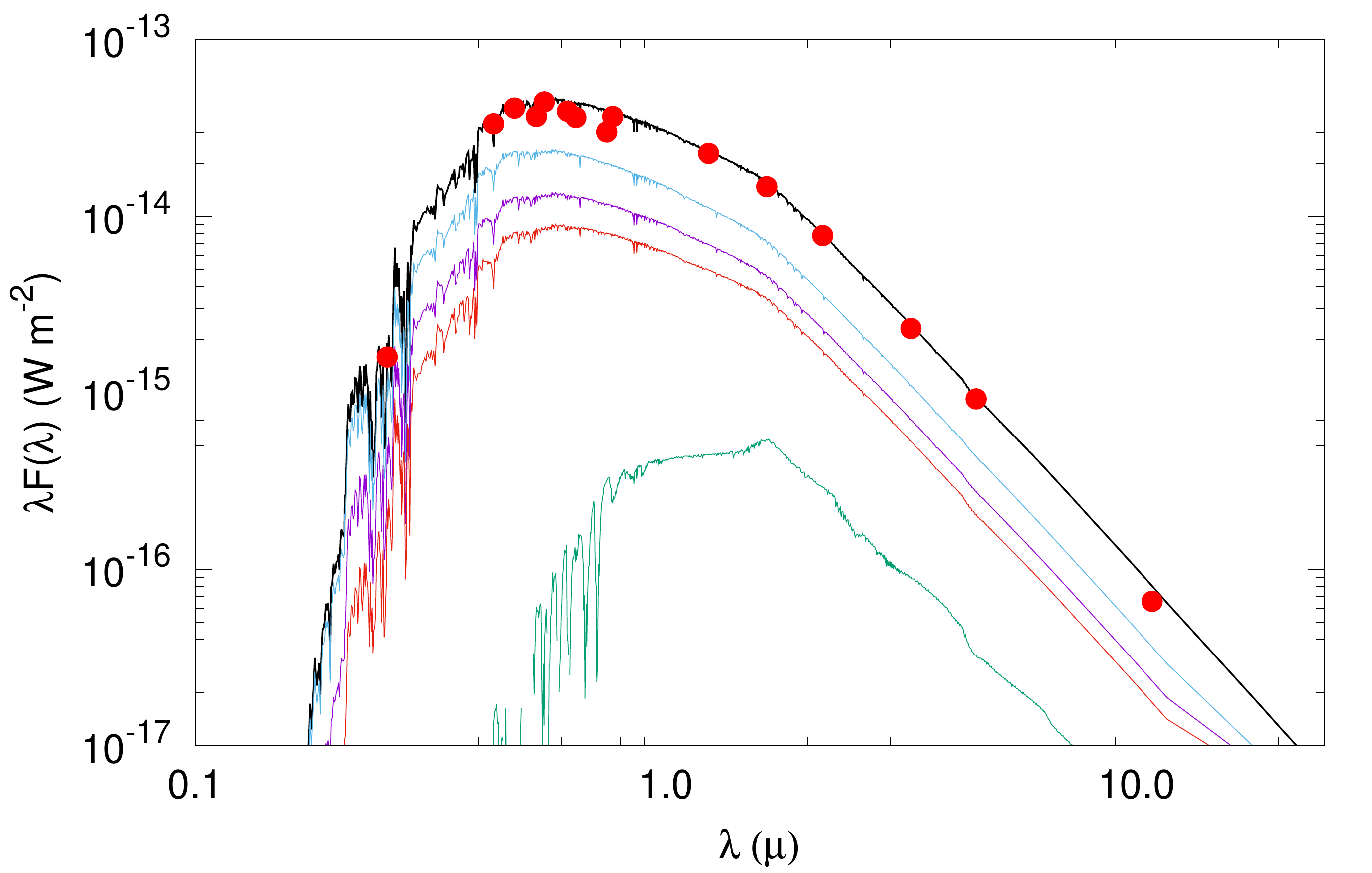}
    \caption{The summed SED of the four stars of TIC\,219006972. The dereddened observed magnitudes are converted into the flux domain (red filled circles), and overplotted with the quasi-continuous summed SED for the quadruple star system (thick black line). This SED is computed from the \citet{2003IAUS..210P.A20C} ATLAS9 stellar atmospheres models (\url{http://www.user.oats.inaf.it/castelli/grids/gridp00k2odfnew/fp00k2tab.html}). The separate SEDs of the four stars (from components Aa, Ab, Ba, and Bb) are also shown with thin cyan, purple, brown and green lines, respectively. } 
   \label{fig:SED}
\end{figure}

\begin{figure}
    \centering
    \includegraphics[width=0.99\linewidth]{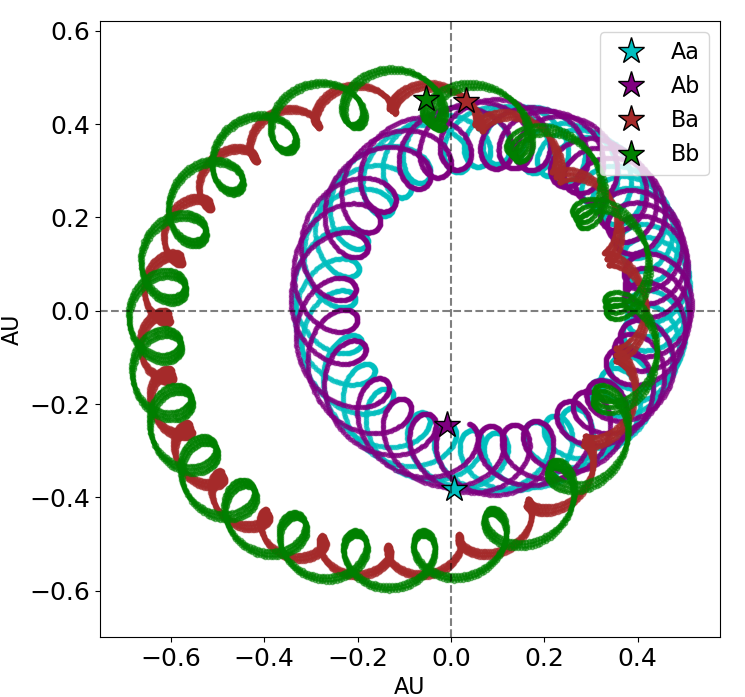}
    \caption{Orbital configuration of TIC 219006972 as seen from above over the course of a few outer periods. The individual components are color-coded as in Fig. \ref{fig:SED}, i.e. cyan, purple, brown and green lines for Aa, Ab, Ba, and Bb, respectively. The observer is in the x-y plane, looking along the x = 0 direction. The orbits appear thick due to slow orbital precession.}
    \label{fig:top_view}
\end{figure}

Table~\ref{tbl:simlightcurve} lists the median values of the stellar and orbital parameters of the quadruple system that are being either adjusted, internally constrained, or derived from the MCMC posteriors, together with the corresponding $1\sigma$ statistical uncertainties. The ETV curves, sections of the lightcurves, and the SED model of the lowest $\chi^2_\mathrm{global}$ solution are plotted in Figs.~\ref{fig:ETVlcfit}, \ref{fig:ETVlcfit2}, and \ref{fig:SED}, respectively. The orbital configuration of the system, as seen from above the plane of the outer orbit, is shown in Figure \ref{fig:top_view} for a few outer periods. 

\begin{table*}
\centering
\caption{Median values of the parameters from the double EB simultaneous lightcurve,  double ETV, joint SED and \texttt{PARSEC} evolutionary track solution from {\sc Lightcurvefactory.}}
\begin{tabular}{lccccc}
\hline
\multicolumn{6}{c}{Orbital elements$^a$} \\
\hline
   & \multicolumn{3}{c}{subsystem}  \\
   & \multicolumn{2}{c}{A} & \multicolumn{2}{c}{B} & A--B \\
  \hline
$P_\mathrm{a}$ [days]            & \multicolumn{2}{c}{$13.6389_{-0.0024}^{+0.0025}$} & \multicolumn{2}{c}{$8.23247_{-0.00049}^{+0.00050}$}       & $168.187_{-0.071}^{+0.069}$ \\
semimajor axis  [$R_\odot$]      & \multicolumn{2}{c}{$29.69_{-0.66}^{+0.67}$}          & \multicolumn{2}{c}{$18.84_{-0.33}^{+0.38}$}                & $189.2_{-3.9}^{+4.2}$ \\  
$i$ [deg]                        & \multicolumn{2}{c}{$89.45_{-0.12}^{+0.20}$}          & \multicolumn{2}{c}{$89.53_{-0.39}^{+0.32}$}                & $88.52_{-0.45}^{+0.38}$  \\
$e$                              & \multicolumn{2}{c}{$0.01518_{-0.00026}^{+0.00026}$}  & \multicolumn{2}{c}{$0.01847_{-0.00066}^{+0.00070}$}        & $0.2543_{-0.0023}^{+0.0022}$ \\  
$\omega$ [deg]                   & \multicolumn{2}{c}{$303.40_{-0.72}^{+0.70}$}         & \multicolumn{2}{c}{$260.09_{-0.99}^{+1.00}$}               & $16.19_{-0.73}^{+0.74}$ \\
$\tau^b$ [BJD]     & \multicolumn{2}{c}{$2\,458\,673.8396_{-0.0269}^{+0.0259}$}           & \multicolumn{2}{c}{$2\,458\,689.0956_{-0.0221}^{+0.0223}$} & $2\,458\,639.250_{-0.409}^{+0.415}$\\
$\Omega$ [deg]                   & \multicolumn{2}{c}{$0.0$}          & \multicolumn{2}{c}{$-0.26_{-0.67}^{+0.98}$}                                  & $-0.62_{-0.35}^{+0.28}$ \\
$(i_\mathrm{m})_{A-...}^c$ [deg]   & \multicolumn{2}{c}{$0.0$}                            & \multicolumn{2}{c}{$0.80_{-0.33}^{+0.39}$}                 & $1.12_{-0.45}^{+0.64}$ \\
$(i_\mathrm{m})_{B-...}$ [deg]   & \multicolumn{2}{c}{$0.80_{-0.33}^{+0.39}$}           & \multicolumn{2}{c}{$0.0$}                                  & $1.31_{-0.47}^{+0.72}$ \\
$\varpi_\mathrm{dyn}^d$ [deg      ]& \multicolumn{2}{c}{$123.40_{-0.72}^{+0.70}$} & \multicolumn{2}{c}{$80.10_{-0.99}^{+1.00}$} & $196.18_{-0.73}^{+0.74}$ \\
$i_\mathrm{dyn}^d$ [deg] & \multicolumn{2}{c}{$0.88_{-0.37}^{+0.53}$} & \multicolumn{2}{c}{$1.08_{-0.42}^{+0.60}$} & $0.25_{-0.09}^{+0.12}$ \\
$\Omega_\mathrm{dyn}^d$ [deg] & \multicolumn{2}{c}{$213.70_{-7.00}^{+4.43}$} & \multicolumn{2}{c}{$202.31_{-41.98}^{+35.04}$} & $30.27_{-11.31}^{+11.89}$ \\
$i_\mathrm{inv}^e$ [deg] & \multicolumn{5}{c}{$88.72_{-0.35}^{+0.31}$} \\
$\Omega_\mathrm{inv}^e$ [deg] & \multicolumn{5}{c}{$-0.49_{-0.29}^{+0.24}$} \\
mass ratio $[q=m_\mathrm{sec}/m_\mathrm{pri}]$ & \multicolumn{2}{c}{$0.931_{-0.005}^{+0.005}$} & \multicolumn{2}{c}{$0.563_{-0.026}^{+0.023}$} & $0.705_{-0.010}^{+0.012}$ \\
$K_\mathrm{pri}$ [km\,s$^{-1}$] & \multicolumn{2}{c}{$53.1_{-1.1}^{+1.1}$} & \multicolumn{2}{c}{$41.6_{-1.2}^{+1.4}$} & $24.3_{-0.4}^{+0.5}$ \\ 
$K_\mathrm{sec}$ [km\,s$^{-1}$] & \multicolumn{2}{c}{$57.1_{-1.4}^{+1.3}$} & \multicolumn{2}{c}{$74.1_{-1.6}^{+2.0}$} & $34.6_{-0.9}^{+0.7}$ \\ 
  \hline  
\multicolumn{6}{c}{\textbf{Apsidal and nodal motion related parameters$^f$}} \\  
\hline  
$P_\mathrm{apse}$ [year] & \multicolumn{2}{c}{$16.59_{-0.13}^{+0.13}$} & \multicolumn{2}{c}{$19.35_{-0.14}^{+0.17}$} & $59.54_{-0.44}^{+0.42}$\\
$P_\mathrm{apse}^\mathrm{dyn}$ [year] & \multicolumn{2}{c}{$7.57_{-0.05}^{+0.05}$} & \multicolumn{2}{c}{$9.29_{-0.06}^{+0.07}$} & $13.74_{-0.09}^{+0.10}$\\
$P_\mathrm{node}^\mathrm{dyn}$ [year] & \multicolumn{2}{c}{$13.92_{-0.07}^{+0.08}$} & \multicolumn{2}{c}{$17.86_{-0.13}^{+0.12}$} & \\
$\Delta\omega_\mathrm{3b}^\mathrm{dyn}$ [arcsec/cycle] & \multicolumn{2}{c}{$6391_{-40}^{+45}$} & \multicolumn{2}{c}{$3144_{-25}^{+21}$} & $43426_{-303}^{+284}$\\
$\Delta\omega_\mathrm{GR}$ [arcsec/cycle] & \multicolumn{2}{c}{$0.53_{-0.02}^{+0.02}$} & \multicolumn{2}{c}{$0.58_{-0.02}^{+0.02}$} & $0.150_{-0.006}^{+0.007}$\\
$\Delta\omega_\mathrm{tide}$ [arcsec/cycle] & \multicolumn{2}{c}{$0.097_{-0.003}^{+0.003}$} & \multicolumn{2}{c}{$0.103_{-0.011}^{+0.017}$} & $-$\\
\hline
\multicolumn{6}{c}{Stellar parameters} \\
\hline
   & Aa & Ab &  Ba & Bb & \\
  \hline
 \multicolumn{6}{c}{Relative quantities} \\
  \hline
fractional radius [$R/a$]               & $0.0380_{-0.0004}^{+0.0004}$ & $0.0318_{-0.0005}^{+0.0005}$  & $0.0441_{-0.0012}^{+0.0018}$ & $0.0246_{-0.0010}^{+0.0014}$ & \\
fractional flux [in \textit{TESS}-band] & $0.468_{-0.023}^{+0.019}$    & $0.287_{-0.008}^{+0.012}$     & $0.195_{-0.016}^{+0.021}$    & $0.011_{-0.001}^{+0.002}$    & \\
 \hline
 \multicolumn{6}{c}{Physical Quantities} \\
  \hline 
 $m$ [M$_\odot$]   & $0.977_{-0.065}^{+0.069}$ & $0.908_{-0.058}^{+0.061}$ & $0.846_{-0.045}^{+0.054}$ & $0.475_{-0.024}^{+0.032}$ & \\
 $R^g$ [R$_\odot$] & $1.128_{-0.022}^{+0.020}$ & $0.943_{-0.035}^{+0.034}$ & $0.831_{-0.031}^{+0.047}$ & $0.464_{-0.025}^{+0.033}$ & \\
 $T_\mathrm{eff}^g$ [K]& $6162_{-100}^{+85}$   & $5942_{-94}^{+75}$        & $5676_{-95}^{+123}$        & $3698_{-45}^{+51}$        & \\
 $L_\mathrm{bol}^g$ [L$_\odot$] & $1.643_{-0.149}^{+0.154}$ & $0.993_{-0.124}^{+0.131}$ & $0.642_{-0.083}^{+0.139}$ & $0.036_{-0.005}^{+0.008}$ &\\
 $M_\mathrm{bol}^g$ & $4.23_{-0.10}^{+0.10}$    & $4.78_{-0.13}^{+0.14}$    & $5.25_{-0.21}^{+0.15}$    & $8.38_{-0.21}^{+0.15}$    &\\
 $M_V^g           $ & $4.26_{-0.11}^{+0.11}$    & $4.83_{-0.15}^{+0.16}$    & $5.34_{-0.23}^{+0.17}$    & $9.70_{-0.25}^{+0.19}$    &\\
 $\log g^g$ [dex]   & $4.322_{-0.018}^{+0.015}$ & $4.446_{-0.007}^{+0.007}$ & $4.523_{-0.023}^{+0.018}$ & $4.780_{-0.032}^{+0.025}$ &\\
 \hline
\multicolumn{6}{c}{Global Quantities} \\
\hline
$\log$(age)$^g$ [dex] &\multicolumn{5}{c}{$9.779_{-0.172}^{+0.142}$} \\
$ [M/H]^g$  [dex]      &\multicolumn{5}{c}{$-0.275_{-0.082}^{+0.070}$} \\
$E(B-V)$ [mag]    &\multicolumn{5}{c}{$0.048_{-0.026}^{+0.020}$} \\
$(M_V)_\mathrm{tot}^g$           &\multicolumn{5}{c}{$3.52_{-0.13}^{+0.13}$} \\
distance [pc]                &\multicolumn{5}{c}{$1327_{-45}^{+50}$}  \\  
\hline
\end{tabular}
\label{tbl:simlightcurve}

{\em Notes.} (a) Instantaneous, osculating orbital elements at epoch $t_0=2\,458325.0$; (b) Time of periastron passsage; (c) Mutual (relative) inclination; (d) Longitude of pericenter ($\varpi_\mathrm{dyn}$) and inclination ($i_\mathrm{dyn}$) with respect to the dynamical (relative) reference frame (see text for details); (e) Inclination ($i_\mathrm{inv}$) and node ($\Omega_\mathrm{inv}$) of the invariable plane to the sky; (f) See Sect~\ref{Sect:system_properties} for a detailed discussion of the tabulated apsidal motion parameters; (g) Interpolated from the \texttt{PARSEC} isochrones;  
\end{table*}

\section{Discussion}

\subsection{System properties}
\label{Sect:system_properties}

According to the results of our photodynamical analysis of the TIC 219006972 system, the longer period binary A is formed by two similar Sun-like stars with masses $m_\mathrm{Aa}=0.98\pm0.07\,\mathrm{M}_\odot$ and $m_\mathrm{Ab}=0.91\pm0.06\,\mathrm{M}_\odot$, respectively. We note that the relatively large (6-7\%) uncertainties in the masses are a consequence of the lack of RV measurements. In the absence of such data, the absolute masses of the stars are constrained only by (i) the dynamical perturbations through the eclipse timings and, (ii) the combinations of \texttt{PARSEC} isochrones and cumulative SED fittings, as described and discussed in detail in \citet{2022MNRAS.510.1352B}. The mass ratio of binary A, however, is much better constrained to $q_\mathrm{A}=0.931\pm0.005$. In contrast, the shorter period binary B consists of two less massive, largely unequal mass stars ($q_\mathrm{B}=0.56\pm0.02$). The primary in binary B is a $m_\mathrm{Ba}=0.85\pm0.05\,\mathrm{M}_\odot$ late G-type star, and the secondary is a smaller, $m_\mathrm{Bb}=0.48\pm0.03\,\mathrm{M}_\odot$ red dwarf. The best-fit solution implies that the quadruple system is most likely moderately metal deficient relative to the Sun $[M/H]=-0.28\pm0.08$, and has an age of $\tau=6\pm2$\,Gyr-old.

The most remarkable property of TIC 219006972 is its very tight and compact orbital configuration. Hence, in what follows, we investigate this property in detail. Specifically, we discuss the current dynamical and observational implications of the system's configuration and evaluate its long-term dynamical stability. 

According to our results the system is quite flat -- none of the three mutual inclination angles amongst orbital planes A, B and AB exceeds 2\degr (within the corresponding $1\sigma$ uncertainties). These findings are in accord with the fact that the deeper-eclipses of binary A were also detected in the ASAS-SN and ATLAS data (see Fig.~\ref{fig:vetting}). Thus, one can exclude the possibility of large amplitude orbital plane precession and the corresponding rapid disappearance of the eclipses. The characteristic nodal precession timescale of $P_\mathrm{node}=13.9\pm0.1$\,yr (see also Table~\ref{tbl:simlightcurve}), would cause the eclipse depths to change noticeably even over the decade spanned by the ATLAS and ASAS-SN data.

While we cannot detect precession of the orbital plane, the strong dynamical perturbations manifest themselves via at least two prominent observable effects. First, the ETV curves of both binaries exhibit 1-2\,h amplitude quasi-sinusoidal variations with the outer period of $P_\mathrm{out}=168$\,d (upper panels of Fig.~\ref{fig:ETVlcfit}). These quasi-sinusoidal cycles, which are in phase for both binaries and operate on a medium ($P_\mathrm{out}$)-timescale, are manifestations of the perturbations of each binary due to the Keplerian motion of the other \citep[see, e.g.][]{2015MNRAS.448..946B}. We note that these should not be confused with the purely geometrical LTTE, which follows a similar period but produces strictly anticorrelated ETVs for the two binaries, and has only a minor contribution to the measured ETVs\footnote{The LTTE is fully taken into account by the photodynamical model.}. 

The other prominent perturbation effect is the rapid, dynamically-forced apsidal motion, which manifests itself in the divergent behaviour of the primary (red) and secondary (blue) ETV curves. Table~\ref{tbl:simlightcurve} provides the theoretical apsidal motion periods both in the observational and the dynamical frame of reference\footnote{The calculations of these periods and the difference between the observational and dynamical reference system are explained briefly in, e.g., \citet{2021ApJ...917...93K}.}. The predicted apsidal motion periods in the observational frame of reference for the two binaries are $P_\mathrm{apsidal,A}=16.6\pm0.1$\,yr and $P_\mathrm{apsidal,B}=19.4\pm0.2$\,yr. Besides these, we also list the cycle by cycle angular contribution of the classic tidal, general relativistic and the dynamical third (fourth) body effects to the advance of the (dynamical) argument of pericenter\footnote{For an explanation of these terms, see again \citet{2021ApJ...917...93K}.}. As seen from the table, the effect of the dynamical perturbations of each binary to the other completely dominates the relativistic and the classic tidal apsidal motion effects.

\begin{figure*}
    \centering
    \includegraphics[width=0.75\linewidth]{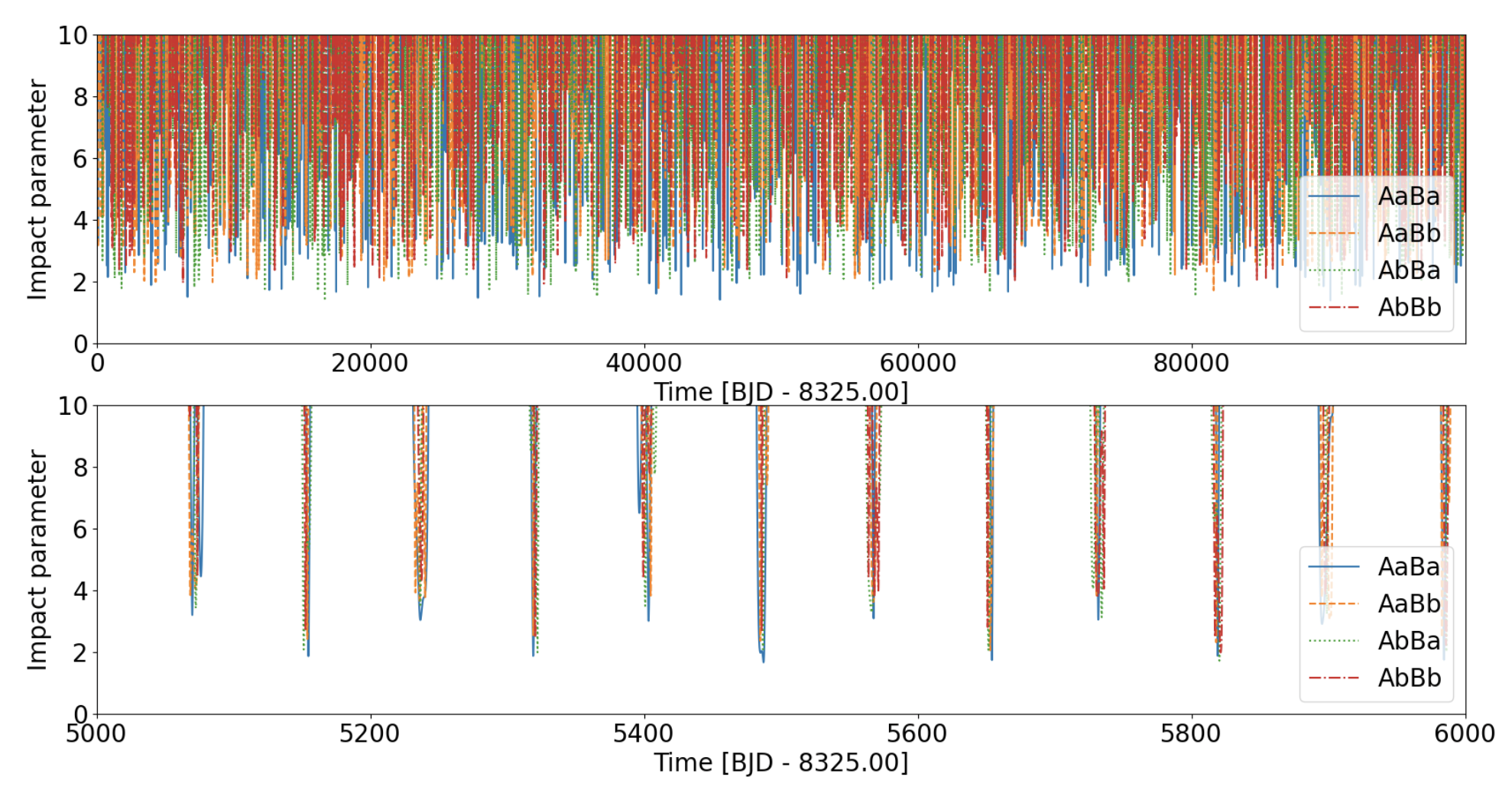}
    \caption{Upper panel: Evolution of the separation between stellar centers of each star of one binary with radius $R_i$, with respect to each star of the other binary with radius $R_j$ in units of $R_i+R_j$ (i.e., the `impact parameters') over the course of 100,000 days. For example, AaBb represents the impact parameter between the primary star of binary A with respect to the secondary star of binary B. When the impact parameter is $< 1$ there is a corresponding eclipse. Lower panel: same as upper panel but zoomed-in on a period of 1,000 days. }
    \label{fig:impact_param}
\end{figure*}

Similar to the two eclipsing pairs, the outer orbital plane was also found to be very close to a perfectly edge-on view ($i_\mathrm{out}=88.5\degr\pm0.4\degr$), raising the interesting question whether TIC 219006972 produces mutual eclipses. Using the best-fit parameters and numerical simulations, we calculated the impact parameters between each star from each binary with respect to each star from the other binary. As demonstrated in Figure \ref{fig:impact_param}, our results show that the mutual impact parameters do not reach below 2 for the duration of the integrations, indicating that the two binaries barely miss each other on-sky and suggesting that mutual inclinations are unlikely in the near future. 

\subsection{Origin}

Multiple stars form through two main channels: (i) fragmentation of cores and filaments of molecular clouds on large (0.01--0.1 pc) scales and (ii) fragmentation of gravitationally unstable protostellar disks at smaller (10--500 AU) separations \citep[see][for a review]{2022arXiv220310066O}. Dynamical friction, disk capture, and circumbinary disk accretion can drive the binary orbits to shorter separations. For example, compact coplanar triples, such as the majority of Kepler EBs exhibiting eclipse timing variations \citep{2016MNRAS.455.4136B}, likely formed via two successive episodes of disk fragmentation and inward disk migration \citep{2016Natur.538..483T,2020MNRAS.491.5158T}. However, not all compact triples survive the embedded protostellar stage as some inner binaries may migrate too far in and merge, and some outer tertiaries may migrate into the dynamical instability regime, leading to an ejection of one of the components. 

Although multiple episodes of disk fragmentation and migration can create compact triples and even (2+1)+1 quadruples, disk fragmentation alone cannot readily produce 2+2 quadruples. Instead, compact 2+2 quadruples likely first derived from core fragmentation on wider scales. Both resulting cores subsequently collapsed into protostellar disks and fragmented again to create the inner companions. Disk fragmentation and inward disk migration can harden binaries down to $P$ $\approx$ 1 day \citep{2018ApJ...854...44M,2020MNRAS.491.5158T}, but the processes of core fragmentation and subsequent dynamical friction tend to leave binaries in slightly larger orbits. Hence, the most compact triple, $\lambda$ Tau at $P_{\rm out}$ = 33 days, is substantially tighter than the most compact 2+2 quadruple presented here, TIC 219006972 with $P_{\rm out}$ = 168 days.

In their toy model of disk fragmentation, accretion, and migration, \cite{2020MNRAS.491.5158T} successfully reproduced the short-end tail of the solar-type binary period distribution (see their Fig. 4). Beyond $P$ $>$ 100 days, however, their simulated population underpredicted the observed rate, suggesting other formation mechanisms, e.g., core fragmentation begins to contribute at these wider separations. This newly discovered compact quadruple is just beyond the expected transition.  Moreover, TIC 219006972 provides some of the strongest evidence that solar-type multiples that initially formed via core fragmentation on $\sim$\,10,000 AU scales can migrate all the way down to $\sim$\,1 AU. 

\subsection{Dynamical Stability and Orbital Evolution}

Given there are two binary stars with a total mass of about 3 ${M_\odot}$ orbiting each other within less than 1 AU (0.877 AU at pericenter), it is quite interesting and instructive to evaluate the long-term stability and orbital evolution of the system. Moreover, we point out that this system is quite `tight', i.e., with small ratios of $P_{\rm out}/P_{\rm in} \simeq 12.3$ and 20.4 for the A and B binaries, respectively.  To check on the system stability, we first use the analytic fitting formalae of \cite{2001MNRAS.321..398M} to calculate the critical semi-major axis and period of the quadruple (below which the system would tend to be dynamically unstable). From Eqn.~(16) of \cite{2013ApJ...768...33R}, the latter must be ${\rm P_{quadruple} > 135~days}$. This is formally below the actual period of 168 days, suggesting that the system is indeed long-term stable and worthy of numerical investigations.

\begin{figure*}
    \centering
    \includegraphics[width=0.85\linewidth]{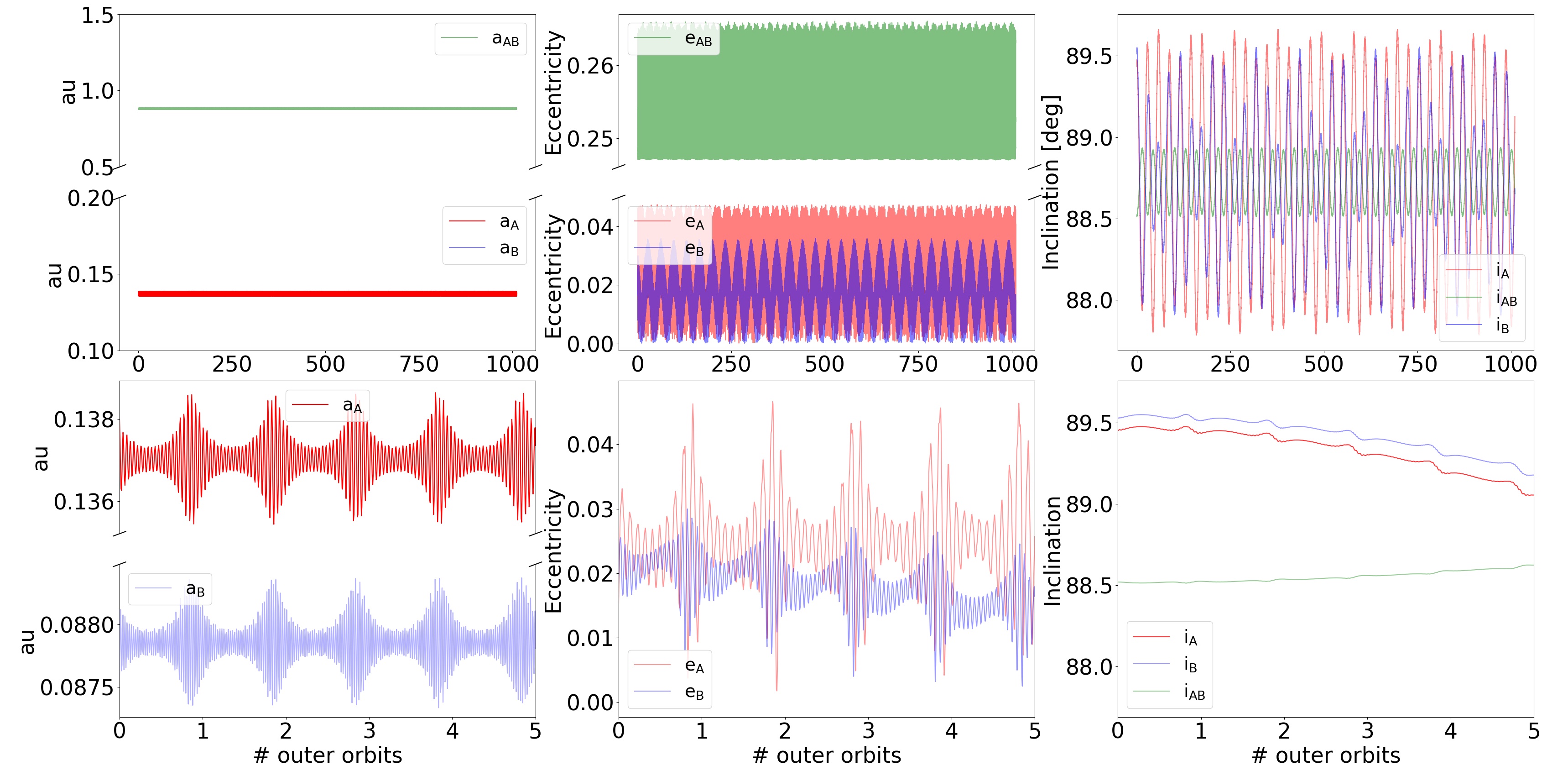}
    \caption{Dynamical evolution of the system shows no signs of chaotic motion. The first column represents the evolution of the three semi-major axes as a function of the outer orbit number, the second column represents the evolution of the three orbital eccentricities and the third columns represents the evolution of the three inclinations. The upper row shows the evolution over the course of 1,000 outer orbits ($\approx$168,000 days) and the lower row shows the evolution over the course of 5 outer orbits. The strong effect of the outer periastron passage on the orbital elements of the two binaries is clearly seen in the lower panels.}%
    \label{fig:dynamics_170k}
\end{figure*}

\begin{figure*}
    \centering
    \includegraphics[width=0.75\linewidth]{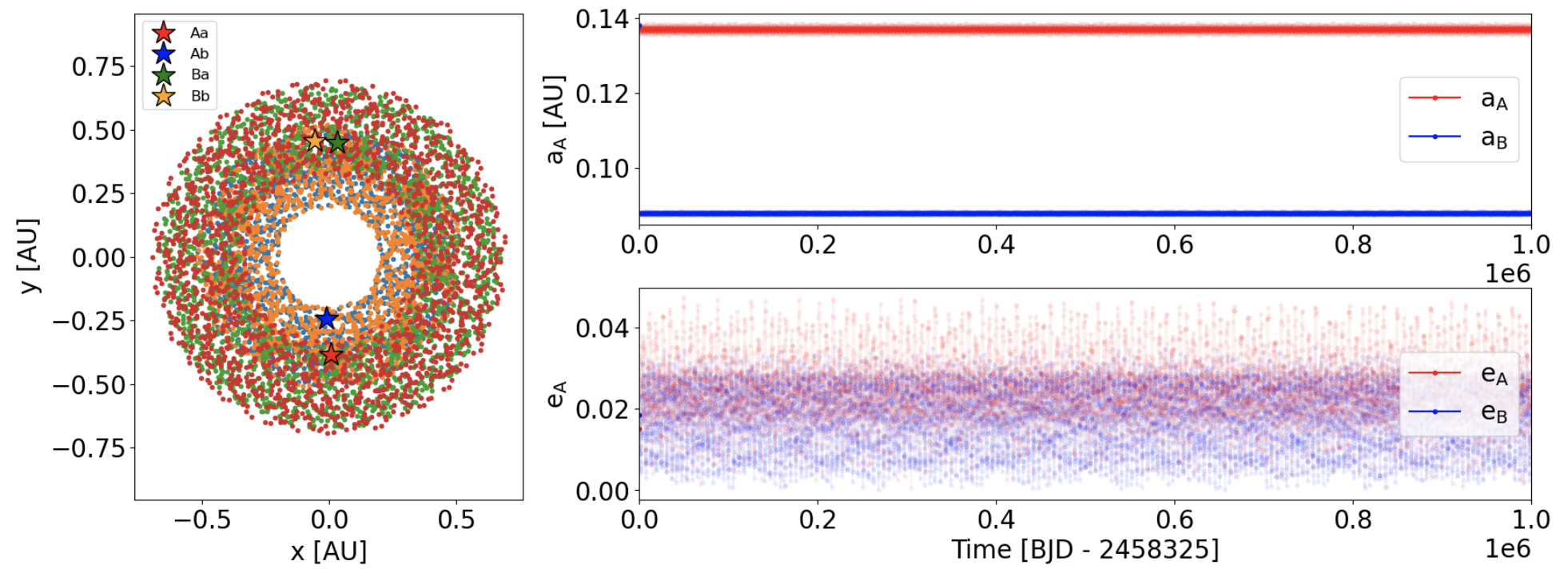}
    \caption{Numerical integrations of the system for 1 million days using the best-fit photodynamical parameters. Left: Orbital configuration of the system. Right panels: Evolution of the semi-major axes (upper) and eccentricities (lower panels) of the two binaries. The system remains stable for the duration of the integrations.}
    \label{fig:dynamics_1m}
\end{figure*}

Therefore, we performed N-body simulations based on the best-fit parameters from the photodynamical solution using the REBOUND package \citep{Rein12}. An example from these simulations is shown in Figure \ref{fig:dynamics_170k} for the case of dynamical integrations spanning 168,000 days ($\approx$1,000 outer orbits). As seen from the figure, the orbital elements of each of the two binaries and the quadruple itself oscillate around their respective mean values without any indications of chaotic behavior for the duration of the integrations. As highlighted in Figure \ref{fig:dynamics_1m}, numerical simulations extending to 1 million days (about 6,000 outer orbits) confirm that the system to exhibit the same behavior as seen in the shorter integration. The semi-major axes of the two binaries oscillate by no more than 2.5\% and the corresponding eccentricities reach no higher than about 0.05; the semi-major of the quadruple varies between 0.877 AU and 0.889 AU, and its eccentricity varies between 0.247 and 0.266. 

We note that since our numerical simulations cover only a tiny fraction of the system's lifetime, it would be interesting to integrate the TIC 219006972 system for a substantial amount of time ($10^6-10^9$ years) and probe its long-term dynamics.  This would be instructive to do since the system is somewhat close to its dynamical stability limit.  Of course, the bottom line is that at least we know the system is empirically stable on Gyr timescales since its age is $6 \pm 2$ Gyr old.

\section{Summary}

In this work, we have presented the discovery of TIC 219006972 -- a compact, eclipsing, nearly co-planar quadruple star producing two sets of primary and secondary eclipses in {\em TESS} data. The two eclipsing binaries have orbital periods of PA = 13.7 days and PB = 8.3 days, respectively, rather small eccentricities (about 0.02), and orbit each other every 168 days on a slightly eccentric outer orbit ($e_\mathrm{out} = 0.25$). This makes TIC 219006972 the shortest-period quadruple system reported to date, and the second closest to co-planarity. 

The masses of the component stars range from 0.48 $M_\odot$ to 0.98 $M_\odot$, their radii from 0.46 $R_\odot$ to 1.13 $R_\odot$, and the respective effective temperatures from 3698 K to 6162 K. The systems is slightly metal-deficient ([M/H] = -0.28), has an age of about 6 Gyr, and appears empirically to be long-term dynamically stable -- in agreement with the analytic stability formalism of \citet{2001MNRAS.321..398M}.

TIC 219006972 raises some intriguing questions such as (i) are there quadruples with substantially shorter outer periods, e.g., $\lesssim 100$ days, and (ii) is there a fundamental lower limit below which quadruples simply cannot form? With the continued
operation of the {\it TESS} and Gaia missions, we hope to further explore these questions.

\section{Data Availability}

The data underlying this article will be shared on reasonable request to the corresponding author.

\clearpage
\begin{acknowledgments}

This paper includes data collected by the {\em TESS} mission, which are publicly available from the Mikulski Archive for Space Telescopes (MAST). Funding for the {\em TESS} mission is provided by NASA's Science Mission directorate. 

This research has made use of the Exoplanet Follow-up Observation Program website, which is operated by the California Institute of Technology, under contract with the National Aeronautics and Space Administration under the Exoplanet Exploration Program. 

Resources supporting this work were provided by the NASA High-End Computing (HEC) Program through the NASA Center for Climate Simulation (NCCS) at Goddard Space Flight Center. Personnel directly supporting this effort were Mark L. Carroll, Laura E. Carriere, Ellen M. Salmon, Nicko D. Acks, Matthew J. Stroud, Bruce E. Pfaff, Lyn E. Gerner, Timothy M. Burch, and Savannah L. Strong.

This work has made use of data from the European Space Agency (ESA) mission {\it Gaia} (\url{https://www.cosmos.esa.int/gaia}), processed by the {\it Gaia} Data Processing and Analysis Consortium (DPAC, \url{https://www.cosmos.esa.int/web/gaia/dpac/consortium}). Funding for the DPAC has been provided by national institutions, in particular the institutions participating in the {\it Gaia} Multilateral Agreement. This work made use of \texttt{tpfplotter} by J. Lillo-Box (publicly available at \url{www.github.com/jlillo/tpfplotter}), which also made use of the python packages \texttt{astropy}, \texttt{lightkurve}, \texttt{matplotlib} and \texttt{numpy}.

A.\,P. acknowledges the financial support of the Hungarian National Research, Development and Innovation Office -- NKFIH Grant K-138962. V.\,B.\,K. is grateful for financial support from NASA grants 80NSSC21K0631 and 80NSSC22K0190.
\end{acknowledgments}

\facilities{
\emph{Gaia},
MAST,
TESS,
ASAS-SN,
ATLAS,
NCCS,
}

\software{
{\tt Astropy} \citep{astropy2013,astropy2018}, 
{\tt Eleanor} \citep{eleanor},
{\tt IPython} \citep{ipython},
{\tt Keras} \citep{keras},
{\tt Keras-vis} \citep{kerasvis},
{\tt LcTools} \citep{2019arXiv191008034S,2021arXiv210310285S},
{\tt Lightcurvefactory} \citep{Borkovits2019,Borkovits2020},
{\tt Lightkurve} \citep{lightkurve},
{\tt Matplotlib} \citep{matplotlib},
{\tt Mpi4py} \citep{mpi4py2008},
{\tt NumPy} \citep{numpy}, 
{\tt Pandas} \citep{pandas},
{\tt Scikit-learn} \citep{scikit-learn},
{\tt SciPy} \citep{scipy},
{\tt Tensorflow} \citep{tensorflow},
{\tt Tess-point} \citep{tess-point}
}

\bibliography{refs}{}

\begin{thebibliography}{}
\expandafter\ifx\csname natexlab\endcsname\relax\def\natexlab#1{#1}\fi
\providecommand{\url}[1]{\href{#1}{#1}}
\providecommand{\dodoi}[1]{doi:~\href{http://doi.org/#1}{\nolinkurl{#1}}}
\providecommand{\doeprint}[1]{\href{http://ascl.net/#1}{\nolinkurl{http://ascl.net/#1}}}
\providecommand{\doarXiv}[1]{\href{https://arxiv.org/abs/#1}{\nolinkurl{https://arxiv.org/abs/#1}}}

\bibitem[{Abadi {et~al.}(2015)Abadi, Agarwal, Barham, Brevdo, Chen, Citro,
  Corrado, Davis, Dean, Devin, Ghemawat, Goodfellow, Harp, Irving, Isard, Jia,
  Jozefowicz, Kaiser, Kudlur, Levenberg, Man\'{e}, Monga, Moore, Murray, Olah,
  Schuster, Shlens, Steiner, Sutskever, Talwar, Tucker, Vanhoucke, Vasudevan,
  Vi\'{e}gas, Vinyals, Warden, Wattenberg, Wicke, Yu, \& Zheng}]{tensorflow}
Abadi, M., Agarwal, A., Barham, P., {et~al.} 2015, {TensorFlow}: Large-Scale
  Machine Learning on Heterogeneous Systems.
\newblock \url{https://www.tensorflow.org/}

\bibitem[{{Aller} {et~al.}(2020){Aller}, {Lillo-Box}, {Jones}, {Miranda}, \&
  {Barcel{\'o} Forteza}}]{2020A&A...635A.128A}
{Aller}, A., {Lillo-Box}, J., {Jones}, D., {Miranda}, L.~F., \& {Barcel{\'o}
  Forteza}, S. 2020, \aap, 635, A128, \dodoi{10.1051/0004-6361/201937118}

\bibitem[{{Astropy Collaboration} {et~al.}(2013){Astropy Collaboration},
  {Robitaille}, {Tollerud}, {Greenfield}, {Droettboom}, {Bray}, {Aldcroft},
  {Davis}, {Ginsburg}, {Price-Whelan}, {Kerzendorf}, {Conley}, {Crighton},
  {Barbary}, {Muna}, {Ferguson}, {Grollier}, {Parikh}, {Nair}, {Unther},
  {Deil}, {Woillez}, {Conseil}, {Kramer}, {Turner}, {Singer}, {Fox}, {Weaver},
  {Zabalza}, {Edwards}, {Azalee Bostroem}, {Burke}, {Casey}, {Crawford},
  {Dencheva}, {Ely}, {Jenness}, {Labrie}, {Lim}, {Pierfederici}, {Pontzen},
  {Ptak}, {Refsdal}, {Servillat}, \& {Streicher}}]{astropy2013}
{Astropy Collaboration}, {Robitaille}, T.~P., {Tollerud}, E.~J., {et~al.} 2013,
  \aap, 558, A33, \dodoi{10.1051/0004-6361/201322068}

\bibitem[{{Astropy Collaboration} {et~al.}(2018){Astropy Collaboration},
  {Price-Whelan}, {Sip{\H{o}}cz}, {G{\"u}nther}, {Lim}, {Crawford}, {Conseil},
  {Shupe}, {Craig}, {Dencheva}, {Ginsburg}, {Vand erPlas}, {Bradley},
  {P{\'e}rez-Su{\'a}rez}, {de Val-Borro}, {Aldcroft}, {Cruz}, {Robitaille},
  {Tollerud}, {Ardelean}, {Babej}, {Bach}, {Bachetti}, {Bakanov}, {Bamford},
  {Barentsen}, {Barmby}, {Baumbach}, {Berry}, {Biscani}, {Boquien}, {Bostroem},
  {Bouma}, {Brammer}, {Bray}, {Breytenbach}, {Buddelmeijer}, {Burke},
  {Calderone}, {Cano Rodr{\'\i}guez}, {Cara}, {Cardoso}, {Cheedella}, {Copin},
  {Corrales}, {Crichton}, {D'Avella}, {Deil}, {Depagne}, {Dietrich}, {Donath},
  {Droettboom}, {Earl}, {Erben}, {Fabbro}, {Ferreira}, {Finethy}, {Fox},
  {Garrison}, {Gibbons}, {Goldstein}, {Gommers}, {Greco}, {Greenfield},
  {Groener}, {Grollier}, {Hagen}, {Hirst}, {Homeier}, {Horton}, {Hosseinzadeh},
  {Hu}, {Hunkeler}, {Ivezi{\'c}}, {Jain}, {Jenness}, {Kanarek}, {Kendrew},
  {Kern}, {Kerzendorf}, {Khvalko}, {King}, {Kirkby}, {Kulkarni}, {Kumar},
  {Lee}, {Lenz}, {Littlefair}, {Ma}, {Macleod}, {Mastropietro}, {McCully},
  {Montagnac}, {Morris}, {Mueller}, {Mumford}, {Muna}, {Murphy}, {Nelson},
  {Nguyen}, {Ninan}, {N{\"o}the}, {Ogaz}, {Oh}, {Parejko}, {Parley}, {Pascual},
  {Patil}, {Patil}, {Plunkett}, {Prochaska}, {Rastogi}, {Reddy Janga},
  {Sabater}, {Sakurikar}, {Seifert}, {Sherbert}, {Sherwood-Taylor}, {Shih},
  {Sick}, {Silbiger}, {Singanamalla}, {Singer}, {Sladen}, {Sooley},
  {Sornarajah}, {Streicher}, {Teuben}, {Thomas}, {Tremblay}, {Turner},
  {Terr{\'o}n}, {van Kerkwijk}, {de la Vega}, {Watkins}, {Weaver}, {Whitmore},
  {Woillez}, {Zabalza}, \& {Astropy Contributors}}]{astropy2018}
{Astropy Collaboration}, {Price-Whelan}, A.~M., {Sip{\H{o}}cz}, B.~M., {et~al.}
  2018, \aj, 156, 123, \dodoi{10.3847/1538-3881/aabc4f}

\bibitem[{{Borkovits}(2022)}]{2022Galax..10....9B}
{Borkovits}, T. 2022, Galaxies, 10, 9, \dodoi{10.3390/galaxies10010009}

\bibitem[{{Borkovits} {et~al.}(2016){Borkovits}, {Hajdu}, {Sztakovics},
  {Rappaport}, {Levine}, {B{\'\i}r{\'o}}, \& {Klagyivik}}]{2016MNRAS.455.4136B}
{Borkovits}, T., {Hajdu}, T., {Sztakovics}, J., {et~al.} 2016, \mnras, 455,
  4136, \dodoi{10.1093/mnras/stv2530}

\bibitem[{{Borkovits} {et~al.}(2015){Borkovits}, {Rappaport}, {Hajdu}, \&
  {Sztakovics}}]{2015MNRAS.448..946B}
{Borkovits}, T., {Rappaport}, S., {Hajdu}, T., \& {Sztakovics}, J. 2015,
  \mnras, 448, 946, \dodoi{10.1093/mnras/stv015}

\bibitem[{{Borkovits} {et~al.}(2020){Borkovits}, {Rappaport}, {Hajdu},
  {Maxted}, {P{\'a}l}, {Forg{\'a}cs-Dajka}, {Klagyivik}, \&
  {Mitnyan}}]{Borkovits2020}
{Borkovits}, T., {Rappaport}, S.~A., {Hajdu}, T., {et~al.} 2020, \mnras, 493,
  5005, \dodoi{10.1093/mnras/staa495}

\bibitem[{{Borkovits} {et~al.}(2019){Borkovits}, {Rappaport}, {Kaye},
  {Isaacson}, {Vanderburg}, {Howard}, {Kristiansen}, {Omohundro},
  {Schwengeler}, {Terentev}, {Shporer}, {Relles}, {Villanueva}, {Tan},
  {Col{\'o}n}, {Blex}, {Haas}, {Cochran}, \& {Endl}}]{Borkovits2019}
{Borkovits}, T., {Rappaport}, S., {Kaye}, T., {et~al.} 2019, \mnras, 483, 1934,
  \dodoi{10.1093/mnras/sty3157}

\bibitem[{{Borkovits} {et~al.}(2022){Borkovits}, {Mitnyan}, {Rappaport},
  {Pribulla}, {Powell}, {Kostov}, {B{\'\i}r{\'o}}, {Cs{\'a}nyi}, {Garai},
  {Gary}, {Kaye}, {Kom{\v{z}}{\'\i}k}, {Terentev}, {Omohundro}, {Gagliano},
  {Jacobs}, {Kristiansen}, {LaCourse}, {Schwengeler}, {Czavalinga}, {Seli},
  {Huang}, {P{\'a}l}, {Vanderburg}, {Rodriguez}, \&
  {Stevens}}]{2022MNRAS.510.1352B}
{Borkovits}, T., {Mitnyan}, T., {Rappaport}, S.~A., {et~al.} 2022, \mnras, 510,
  1352, \dodoi{10.1093/mnras/stab3397}

\bibitem[{{Burke} {et~al.}(2020){Burke}, {Levine}, {Fausnaugh}, {Vanderspek},
  {Barclay}, {Libby-Roberts}, {Morris}, {Sipocz}, {Owens}, {Feinstein}, \&
  {Camacho}}]{tess-point}
{Burke}, C.~J., {Levine}, A., {Fausnaugh}, M., {et~al.} 2020, {TESS-Point: High
  precision TESS pointing tool}, Astrophysics Source Code Library.
\newblock \doeprint{2003.001}

\bibitem[{{Castelli} \& {Kurucz}(2003)}]{2003IAUS..210P.A20C}
{Castelli}, F., \& {Kurucz}, R.~L. 2003, in Modelling of Stellar Atmospheres,
  ed. N.~{Piskunov}, W.~W. {Weiss}, \& D.~F. {Gray}, Vol. 210, A20,
  \dodoi{10.48550/arXiv.astro-ph/0405087}

\bibitem[{Chollet {et~al.}(2015)}]{keras}
Chollet, F., {et~al.} 2015, Keras, \url{https://keras.io}

\bibitem[{{Cutri} {et~al.}(2012){Cutri}, {Wright}, {Conrow}, {Bauer},
  {Benford}, {Brandenburg}, {Dailey}, {Eisenhardt}, {Evans}, {Fajardo-Acosta},
  {Fowler}, {Gelino}, {Grillmair}, {Harbut}, {Hoffman}, {Jarrett},
  {Kirkpatrick}, {Leisawitz}, {Liu}, {Mainzer}, {Marsh}, {Masci}, {McCallon},
  {Padgett}, {Ressler}, {Royer}, {Skrutskie}, {Stanford}, {Wyatt}, {Tholen},
  {Tsai}, {Wachter}, {Wheelock}, {Yan}, {Alles}, {Beck}, {Grav}, {Masiero},
  {McCollum}, {McGehee}, {Papin}, \& {Wittman}}]{WISE}
{Cutri}, R.~M., {Wright}, E.~L., {Conrow}, T., {et~al.} 2012, {Explanatory
  Supplement to the WISE All-Sky Data Release Products}, Explanatory Supplement
  to the WISE All-Sky Data Release Products

\bibitem[{Dalcin {et~al.}(2008)Dalcin, Paz, Storti, \& D'Elia}]{mpi4py2008}
Dalcin, L., Paz, R., Storti, M., \& D'Elia, J. 2008, Journal of Parallel and
  Distributed Computing, 68, 655,
  \dodoi{http://dx.doi.org/10.1016/j.jpdc.2007.09.005}

\bibitem[{{De Rosa} {et~al.}(2014){De Rosa}, {Patience}, {Wilson}, {Schneider},
  {Wiktorowicz}, {Vigan}, {Marois}, {Song}, {Macintosh}, {Graham}, {Doyon},
  {Bessell}, {Thomas}, \& {Lai}}]{DeRosa2014}
{De Rosa}, R.~J., {Patience}, J., {Wilson}, P.~A., {et~al.} 2014, \mnras, 437,
  1216, \dodoi{10.1093/mnras/stt1932}

\bibitem[{{Ebbighausen} \& {Struve}(1956)}]{1956ApJ...124..507E}
{Ebbighausen}, E.~G., \& {Struve}, O. 1956, \apj, 124, 507,
  \dodoi{10.1086/146254}

\bibitem[{{Fang} {et~al.}(2018){Fang}, {Thompson}, \&
  {Hirata}}]{2018MNRAS.476.4234F}
{Fang}, X., {Thompson}, T.~A., \& {Hirata}, C.~M. 2018, \mnras, 476, 4234,
  \dodoi{10.1093/mnras/sty472}

\bibitem[{{Feinstein} {et~al.}(2019){Feinstein}, {Montet}, {Foreman-Mackey},
  {Bedell}, {Saunders}, {Bean}, {Christiansen}, {Hedges}, {Luger}, {Scolnic},
  \& {Cardoso}}]{eleanor}
{Feinstein}, A.~D., {Montet}, B.~T., {Foreman-Mackey}, D., {et~al.} 2019,
  \pasp, 131, 094502, \dodoi{10.1088/1538-3873/ab291c}

\bibitem[{{Fragione} \& {Kocsis}(2019)}]{2019MNRAS.486.4781F}
{Fragione}, G., \& {Kocsis}, B. 2019, \mnras, 486, 4781,
  \dodoi{10.1093/mnras/stz1175}

\bibitem[{{Gaia Collaboration} {et~al.}(2021){Gaia Collaboration}, {Brown},
  {Vallenari}, {Prusti}, {de Bruijne}, {Babusiaux}, {Biermann}, {Creevey},
  {Evans}, {Eyer}, {Hutton}, {Jansen}, {Jordi}, {Klioner}, {Lammers},
  {Lindegren}, {Luri}, {Mignard}, {Panem}, {Pourbaix}, {Randich}, {Sartoretti},
  {Soubiran}, {Walton}, {Arenou}, {Bailer-Jones}, {Bastian}, {Cropper},
  {Drimmel}, {Katz}, {Lattanzi}, {van Leeuwen}, {Bakker}, {Cacciari},
  {Casta{\~n}eda}, {De Angeli}, {Ducourant}, {Fabricius}, {Fouesneau},
  {Fr{\'e}mat}, {Guerra}, {Guerrier}, {Guiraud}, {Jean-Antoine Piccolo},
  {Masana}, {Messineo}, {Mowlavi}, {Nicolas}, {Nienartowicz}, {Pailler},
  {Panuzzo}, {Riclet}, {Roux}, {Seabroke}, {Sordo}, {Tanga}, {Th{\'e}venin},
  {Gracia-Abril}, {Portell}, {Teyssier}, {Altmann}, {Andrae}, {Bellas-Velidis},
  {Benson}, {Berthier}, {Blomme}, {Brugaletta}, {Burgess}, {Busso}, {Carry},
  {Cellino}, {Cheek}, {Clementini}, {Damerdji}, {Davidson}, {Delchambre},
  {Dell'Oro}, {Fern{\'a}ndez-Hern{\'a}ndez}, {Galluccio}, {Garc{\'\i}a-Lario},
  {Garcia-Reinaldos}, {Gonz{\'a}lez-N{\'u}{\~n}ez}, {Gosset}, {Haigron},
  {Halbwachs}, {Hambly}, {Harrison}, {Hatzidimitriou}, {Heiter},
  {Hern{\'a}ndez}, {Hestroffer}, {Hodgkin}, {Holl}, {Jan{\ss}en}, {Jevardat de
  Fombelle}, {Jordan}, {Krone-Martins}, {Lanzafame}, {L{\"o}ffler}, {Lorca},
  {Manteiga}, {Marchal}, {Marrese}, {Moitinho}, {Mora}, {Muinonen}, {Osborne},
  {Pancino}, {Pauwels}, {Petit}, {Recio-Blanco}, {Richards}, {Riello},
  {Rimoldini}, {Robin}, {Roegiers}, {Rybizki}, {Sarro}, {Siopis}, {Smith},
  {Sozzetti}, {Ulla}, {Utrilla}, {van Leeuwen}, {van Reeven}, {Abbas}, {Abreu
  Aramburu}, {Accart}, {Aerts}, {Aguado}, {Ajaj}, {Altavilla}, {{\'A}lvarez},
  {{\'A}lvarez Cid-Fuentes}, {Alves}, {Anderson}, {Anglada Varela}, {Antoja},
  {Audard}, {Baines}, {Baker}, {Balaguer-N{\'u}{\~n}ez}, {Balbinot}, {Balog},
  {Barache}, {Barbato}, {Barros}, {Barstow}, {Bartolom{\'e}}, {Bassilana},
  {Bauchet}, {Baudesson-Stella}, {Becciani}, {Bellazzini}, {Bernet}, {Bertone},
  {Bianchi}, {Blanco-Cuaresma}, {Boch}, {Bombrun}, {Bossini}, {Bouquillon},
  {Bragaglia}, {Bramante}, {Breedt}, {Bressan}, {Brouillet}, {Bucciarelli},
  {Burlacu}, {Busonero}, {Butkevich}, {Buzzi}, {Caffau}, {Cancelliere},
  {C{\'a}novas}, {Cantat-Gaudin}, {Carballo}, {Carlucci}, {Carnerero},
  {Carrasco}, {Casamiquela}, {Castellani}, {Castro-Ginard}, {Castro Sampol},
  {Chaoul}, {Charlot}, {Chemin}, {Chiavassa}, {Cioni}, {Comoretto}, {Cooper},
  {Cornez}, {Cowell}, {Crifo}, {Crosta}, {Crowley}, {Dafonte}, {Dapergolas},
  {David}, {David}, {de Laverny}, {De Luise}, {De March}, {De Ridder}, {de
  Souza}, {de Teodoro}, {de Torres}, {del Peloso}, {del Pozo}, {Delbo},
  {Delgado}, {Delgado}, {Delisle}, {Di Matteo}, {Diakite}, {Diener},
  {Distefano}, {Dolding}, {Eappachen}, {Edvardsson}, {Enke}, {Esquej}, {Fabre},
  {Fabrizio}, {Faigler}, {Fedorets}, {Fernique}, {Fienga}, {Figueras},
  {Fouron}, {Fragkoudi}, {Fraile}, {Franke}, {Gai}, {Garabato},
  {Garcia-Gutierrez}, {Garc{\'\i}a-Torres}, {Garofalo}, {Gavras}, {Gerlach},
  {Geyer}, {Giacobbe}, {Gilmore}, {Girona}, {Giuffrida}, {Gomel}, {Gomez},
  {Gonzalez-Santamaria}, {Gonz{\'a}lez-Vidal}, {Granvik},
  {Guti{\'e}rrez-S{\'a}nchez}, {Guy}, {Hauser}, {Haywood}, {Helmi}, {Hidalgo},
  {Hilger}, {H{\l}adczuk}, {Hobbs}, {Holland}, {Huckle}, {Jasniewicz},
  {Jonker}, {Juaristi Campillo}, {Julbe}, {Karbevska}, {Kervella}, {Khanna},
  {Kochoska}, {Kontizas}, {Kordopatis}, {Korn}, {Kostrzewa-Rutkowska},
  {Kruszy{\'n}ska}, {Lambert}, {Lanza}, {Lasne}, {Le Campion}, {Le Fustec},
  {Lebreton}, {Lebzelter}, {Leccia}, {Leclerc}, {Lecoeur-Taibi}, {Liao},
  {Licata}, {Lindstr{\o}m}, {Lister}, {Livanou}, {Lobel}, {Madrero Pardo},
  {Managau}, {Mann}, {Marchant}, {Marconi}, {Marcos Santos}, {Marinoni},
  {Marocco}, {Marshall}, {Martin Polo}, {Mart{\'\i}n-Fleitas}, {Masip},
  {Massari}, {Mastrobuono-Battisti}, {Mazeh}, {McMillan}, {Messina},
  {Michalik}, {Millar}, {Mints}, {Molina}, {Molinaro}, {Moln{\'a}r},
  {Montegriffo}, {Mor}, {Morbidelli}, {Morel}, {Morris}, {Mulone}, {Munoz},
  {Muraveva}, {Murphy}, {Musella}, {Noval}, {Ord{\'e}novic}, {Orr{\`u}},
  {Osinde}, {Pagani}, {Pagano}, {Palaversa}, {Palicio}, {Panahi}, {Pawlak},
  {Pe{\~n}alosa Esteller}, {Penttil{\"a}}, {Piersimoni}, {Pineau}, {Plachy},
  {Plum}, {Poggio}, {Poretti}, {Poujoulet}, {Pr{\v{s}}a}, {Pulone}, {Racero},
  {Ragaini}, {Rainer}, {Raiteri}, {Rambaux}, {Ramos}, {Ramos-Lerate}, {Re
  Fiorentin}, {Regibo}, {Reyl{\'e}}, {Ripepi}, {Riva}, {Rixon}, {Robichon},
  {Robin}, {Roelens}, {Rohrbasser}, {Romero-G{\'o}mez}, {Rowell}, {Royer},
  {Rybicki}, {Sadowski}, {Sagrist{\`a} Sell{\'e}s}, {Sahlmann}, {Salgado},
  {Salguero}, {Samaras}, {Sanchez Gimenez}, {Sanna}, {Santove{\~n}a},
  {Sarasso}, {Schultheis}, {Sciacca}, {Segol}, {Segovia}, {S{\'e}gransan},
  {Semeux}, {Shahaf}, {Siddiqui}, {Siebert}, {Siltala}, {Slezak}, {Smart},
  {Solano}, {Solitro}, {Souami}, {Souchay}, {Spagna}, {Spoto}, {Steele},
  {Steidelm{\"u}ller}, {Stephenson}, {S{\"u}veges}, {Szabados}, {Szegedi-Elek},
  {Taris}, {Tauran}, {Taylor}, {Teixeira}, {Thuillot}, {Tonello}, {Torra},
  {Torra}, {Turon}, {Unger}, {Vaillant}, {van Dillen}, {Vanel}, {Vecchiato},
  {Viala}, {Vicente}, {Voutsinas}, {Weiler}, {Wevers}, {Wyrzykowski}, {Yoldas},
  {Yvard}, {Zhao}, {Zorec}, {Zucker}, {Zurbach}, \& {Zwitter}}]{Gaia2021}
{Gaia Collaboration}, {Brown}, A.~G.~A., {Vallenari}, A., {et~al.} 2021, \aap,
  649, A1, \dodoi{10.1051/0004-6361/202039657}

\bibitem[{{Hamers} {et~al.}(2021){Hamers}, {Rantala}, {Neunteufel}, {Preece},
  \& {Vynatheya}}]{2021MNRAS.502.4479H}
{Hamers}, A.~S., {Rantala}, A., {Neunteufel}, P., {Preece}, H., \& {Vynatheya},
  P. 2021, \mnras, 502, 4479, \dodoi{10.1093/mnras/stab287}

\bibitem[{Harris {et~al.}(2020)Harris, Millman, van~der Walt, Gommers,
  Virtanen, Cournapeau, Wieser, Taylor, Berg, Smith, Kern, Picus, Hoyer, van
  Kerkwijk, Brett, Haldane, del R{\'{i}}o, Wiebe, Peterson,
  G{\'{e}}rard-Marchant, Sheppard, Reddy, Weckesser, Abbasi, Gohlke, \&
  Oliphant}]{numpy}
Harris, C.~R., Millman, K.~J., van~der Walt, S.~J., {et~al.} 2020, Nature, 585,
  357, \dodoi{10.1038/s41586-020-2649-2}

\bibitem[{{Heinze} {et~al.}(2018){Heinze}, {Tonry}, {Denneau}, {Flewelling},
  {Stalder}, {Rest}, {Smith}, {Smartt}, \& {Weiland}}]{2018AJ....156..241H}
{Heinze}, A.~N., {Tonry}, J.~L., {Denneau}, L., {et~al.} 2018, \aj, 156, 241,
  \dodoi{10.3847/1538-3881/aae47f}

\bibitem[{Hunter(2007)}]{matplotlib}
Hunter, J.~D. 2007, Computing in science \& engineering, 9, 90

\bibitem[{{Kochanek} {et~al.}(2017){Kochanek}, {Shappee}, {Stanek}, {Holoien},
  {Thompson}, {Prieto}, {Dong}, {Shields}, {Will}, {Britt}, {Perzanowski}, \&
  {Pojma{\'n}ski}}]{2017PASP..129j4502K}
{Kochanek}, C.~S., {Shappee}, B.~J., {Stanek}, K.~Z., {et~al.} 2017, \pasp,
  129, 104502, \dodoi{10.1088/1538-3873/aa80d9}

\bibitem[{{Kostov} {et~al.}(2021){Kostov}, {Powell}, {Torres}, {Borkovits},
  {Rappaport}, {Tokovinin}, {Zasche}, {Anderson}, {Barclay}, {Berlind},
  {Brown}, {Calkins}, {Collins}, {Collins}, {Conti}, {Esquerdo}, {Hellier},
  {Jensen}, {Kamler}, {Kruse}, {Latham}, {Ma{\v{s}}ek}, {Murgas}, {Olmschenk},
  {Orosz}, {P{\'a}l}, {Palle}, {Schwarz}, {Stockdale}, {Tamayo}, {Uhla{\v{r}}},
  {Welsh}, \& {West}}]{2021ApJ...917...93K}
{Kostov}, V.~B., {Powell}, B.~P., {Torres}, G., {et~al.} 2021, \apj, 917, 93,
  \dodoi{10.3847/1538-4357/ac04ad}

\bibitem[{{Kostov} {et~al.}(2022){Kostov}, {Powell}, {Rappaport}, {Borkovits},
  {Gagliano}, {Jacobs}, {Kristiansen}, {LaCourse}, {Omohundro}, {Orosz},
  {Schmitt}, {Schwengeler}, {Terentev}, {Torres}, {Barclay}, {Friedman},
  {Kruse}, {Olmschenk}, {Vanderburg}, \& {Welsh}}]{2022ApJS..259...66K}
{Kostov}, V.~B., {Powell}, B.~P., {Rappaport}, S.~A., {et~al.} 2022, \apjs,
  259, 66, \dodoi{10.3847/1538-4365/ac5458}

\bibitem[{Kotikalapudi \& contributors(2017)}]{kerasvis}
Kotikalapudi, R., \& contributors. 2017, keras-vis,
  \url{https://github.com/raghakot/keras-vis},  GitHub

\bibitem[{{Kozai}(1962)}]{1962AJ.....67..591K}
{Kozai}, Y. 1962, \aj, 67, 591, \dodoi{10.1086/108790}

\bibitem[{{Kristiansen} {et~al.}(2022){Kristiansen}, {Rappaport}, {Vanderburg},
  {Jacobs}, {Martin Schwengeler}, {Gagliano}, {Terentev}, {LaCourse},
  {Omohundro}, {Schmitt}, {Powell}, \& {Kostov}}]{2022PASP..134g4401K}
{Kristiansen}, M. H.~K., {Rappaport}, S.~A., {Vanderburg}, A.~M., {et~al.}
  2022, \pasp, 134, 074401, \dodoi{10.1088/1538-3873/ac6e06}

\bibitem[{{Lidov}(1962)}]{Lidov1962}
{Lidov}, M.~L. 1962, \planss, 9, 719, \dodoi{10.1016/0032-0633(62)90129-0}

\bibitem[{{Lightkurve Collaboration} {et~al.}(2018){Lightkurve Collaboration},
  {Cardoso}, {Hedges}, {Gully-Santiago}, {Saunders}, {Cody}, {Barclay}, {Hall},
  {Sagear}, {Turtelboom}, {Zhang}, {Tzanidakis}, {Mighell}, {Coughlin}, {Bell},
  {Berta-Thompson}, {Williams}, {Dotson}, \& {Barentsen}}]{lightkurve}
{Lightkurve Collaboration}, {Cardoso}, J.~V.~d.~M., {Hedges}, C., {et~al.}
  2018, {Lightkurve: Kepler and TESS time series analysis in Python},
  Astrophysics Source Code Library.
\newblock \doeprint{1812.013}

\bibitem[{{Liu} \& {Lai}(2019)}]{2019MNRAS.483.4060L}
{Liu}, B., \& {Lai}, D. 2019, \mnras, 483, 4060, \dodoi{10.1093/mnras/sty3432}

\bibitem[{{Mardling} \& {Aarseth}(2001)}]{2001MNRAS.321..398M}
{Mardling}, R.~A., \& {Aarseth}, S.~J. 2001, \mnras, 321, 398,
  \dodoi{10.1046/j.1365-8711.2001.03974.x}

\bibitem[{McKinney(2010)}]{pandas}
McKinney, W. 2010, in Proceedings of the 9th Python in Science Conference, ed.
  S.~van~der Walt \& J.~Millman, 51 -- 56

\bibitem[{{Moe} \& {Di Stefano}(2017)}]{Moe2017}
{Moe}, M., \& {Di Stefano}, R. 2017, \apjs, 230, 15,
  \dodoi{10.3847/1538-4365/aa6fb6}

\bibitem[{{Moe} \& {Kratter}(2018)}]{2018ApJ...854...44M}
{Moe}, M., \& {Kratter}, K.~M. 2018, \apj, 854, 44,
  \dodoi{10.3847/1538-4357/aaa6d2}

\bibitem[{{Naoz} \& {Fabrycky}(2014)}]{2014ApJ...793..137N}
{Naoz}, S., \& {Fabrycky}, D.~C. 2014, \apj, 793, 137,
  \dodoi{10.1088/0004-637X/793/2/137}

\bibitem[{{Offner} {et~al.}(2022){Offner}, {Moe}, {Kratter}, {Sadavoy},
  {Jensen}, \& {Tobin}}]{2022arXiv220310066O}
{Offner}, S. S.~R., {Moe}, M., {Kratter}, K.~M., {et~al.} 2022, arXiv e-prints,
  arXiv:2203.10066, \dodoi{10.48550/arXiv.2203.10066}

\bibitem[{{P{\'a}l}(2012)}]{2012MNRAS.421.1825P}
{P{\'a}l}, A. 2012, \mnras, 421, 1825, \dodoi{10.1111/j.1365-2966.2011.19813.x}

\bibitem[{Pedregosa {et~al.}(2011)Pedregosa, Varoquaux, Gramfort, Michel,
  Thirion, Grisel, Blondel, Prettenhofer, Weiss, Dubourg, Vanderplas, Passos,
  Cournapeau, Brucher, Perrot, \& Duchesnay}]{scikit-learn}
Pedregosa, F., Varoquaux, G., Gramfort, A., {et~al.} 2011, Journal of Machine
  Learning Research, 12, 2825

\bibitem[{{Pejcha} {et~al.}(2013){Pejcha}, {Antognini}, {Shappee}, \&
  {Thompson}}]{2013MNRAS.435..943P}
{Pejcha}, O., {Antognini}, J.~M., {Shappee}, B.~J., \& {Thompson}, T.~A. 2013,
  \mnras, 435, 943, \dodoi{10.1093/mnras/stt1281}

\bibitem[{{Perets} \& {Fabrycky}(2009)}]{2009ApJ...697.1048P}
{Perets}, H.~B., \& {Fabrycky}, D.~C. 2009, \apj, 697, 1048,
  \dodoi{10.1088/0004-637X/697/2/1048}

\bibitem[{P\'erez \& Granger(2007)}]{ipython}
P\'erez, F., \& Granger, B.~E. 2007, Computing in Science and Engineering, 9,
  21, \dodoi{10.1109/MCSE.2007.53}

\bibitem[{{Powell} {et~al.}(2021){Powell}, {Kostov}, {Rappaport}, {Borkovits},
  {Zasche}, {Tokovinin}, {Kruse}, {Latham}, {Montet}, {Jensen}, {Jayaraman},
  {Collins}, {Ma{\v{s}}ek}, {Hellier}, {Evans}, {Tan}, {Schlieder}, {Torres},
  {Smale}, {Friedman}, {Barclay}, {Gagliano}, {Quintana}, {Jacobs}, {Gilbert},
  {Kristiansen}, {Col{\'o}n}, {LaCourse}, {Olmschenk}, {Omohundro},
  {Schnittman}, {Schwengeler}, {Barry}, {Terentev}, {Boyd}, {Schmitt}, {Quinn},
  {Vanderburg}, {Palle}, {Armstrong}, {Ricker}, {Vanderspek}, {Seager}, {Winn},
  {Jenkins}, {Caldwell}, {Wohler}, {Shiao}, {Burke}, {Daylan}, \&
  {Villase{\~n}or}}]{2021AJ....161..162P}
{Powell}, B.~P., {Kostov}, V.~B., {Rappaport}, S.~A., {et~al.} 2021, \aj, 161,
  162, \dodoi{10.3847/1538-3881/abddb5}

\bibitem[{{Powell} {et~al.}(2022{\natexlab{a}}){Powell}, {Rappaport},
  {Borkovits}, {Kostov}, {Torres}, {Jayaraman}, {Latham},
  {Ku{\v{c}}{\'a}kov{\'a}}, {Garai}, {Pribulla}, {Vanderburg}, {Kruse},
  {Barclay}, {Olmschenk}, {Kristiansen}, {Gagliano}, {Jacobs}, {LaCourse},
  {Omohundro}, {Schwengeler}, {Terentev}, \& {Schmitt}}]{2022ApJ...938..133P}
{Powell}, B.~P., {Rappaport}, S.~A., {Borkovits}, T., {et~al.}
  2022{\natexlab{a}}, \apj, 938, 133, \dodoi{10.3847/1538-4357/ac8934}

\bibitem[{{Powell} {et~al.}(2022{\natexlab{b}}){Powell}, {Kruse}, {Montet},
  {Feinstein}, {Lewis}, {Foreman-Mackey}, {Barclay}, {Quintana}, {Col{\'o}n},
  {Kostov}, {Boyd}, {Smale}, {Mullally}, {Schlieder}, {Schnittman}, {Carroll},
  {Carriere}, {Salmon}, {Strong}, {Acks}, {Pfaff}, {Gerner}, \&
  {Burch}}]{2022RNAAS...6..111P}
{Powell}, B.~P., {Kruse}, E., {Montet}, B.~T., {et~al.} 2022{\natexlab{b}},
  Research Notes of the American Astronomical Society, 6, 111,
  \dodoi{10.3847/2515-5172/ac74c4}

\bibitem[{{Pribulla} {et~al.}(2008){Pribulla}, {Balu{\v{d}}ansk{\'y}},
  {Dubovsk{\'y}}, {Kudzej}, {Parimucha}, {Siwak}, \&
  {Va{\v{n}}ko}}]{Pribulla2008}
{Pribulla}, T., {Balu{\v{d}}ansk{\'y}}, D., {Dubovsk{\'y}}, P., {et~al.} 2008,
  \mnras, 390, 798, \dodoi{10.1111/j.1365-2966.2008.13781.x}

\bibitem[{{Pribulla} {et~al.}(2020){Pribulla}, {Puha}, {Borkovits}, {Budaj},
  {Garai}, {Guenther}, {Hamb{\'a}lek}, {Kom{\v{z}}{\'\i}k}, {Kundra},
  {Szab{\'o}}, \& {Va{\v{n}}ko}}]{Pribulla2020}
{Pribulla}, T., {Puha}, E., {Borkovits}, T., {et~al.} 2020, \mnras, 494, 178,
  \dodoi{10.1093/mnras/staa699}

\bibitem[{{Raghavan} {et~al.}(2010){Raghavan}, {McAlister}, {Henry}, {Latham},
  {Marcy}, {Mason}, {Gies}, {White}, \& {ten Brummelaar}}]{Raghavan2010}
{Raghavan}, D., {McAlister}, H.~A., {Henry}, T.~J., {et~al.} 2010, \apjs, 190,
  1, \dodoi{10.1088/0067-0049/190/1/1}

\bibitem[{{Rappaport} {et~al.}(2013){Rappaport}, {Deck}, {Levine}, {Borkovits},
  {Carter}, {El Mellah}, {Sanchis-Ojeda}, \& {Kalomeni}}]{2013ApJ...768...33R}
{Rappaport}, S., {Deck}, K., {Levine}, A., {et~al.} 2013, \apj, 768, 33,
  \dodoi{10.1088/0004-637X/768/1/33}

\bibitem[{{Rappaport} {et~al.}(2022){Rappaport}, {Borkovits}, {Gagliano},
  {Jacobs}, {Kostov}, {Powell}, {Terentev}, {Omohundro}, {Torres},
  {Vanderburg}, {Mitnyan}, {Kristiansen}, {LaCourse}, {Schwengeler}, {Kaye},
  {P{\'a}l}, {Pribulla}, {B{\'\i}r{\'o}}, {Cs{\'a}nyi}, {Garai}, {Zasche},
  {Maxted}, {Rodriguez}, \& {Stevens}}]{2022MNRAS.513.4341R}
{Rappaport}, S.~A., {Borkovits}, T., {Gagliano}, R., {et~al.} 2022, \mnras,
  513, 4341, \dodoi{10.1093/mnras/stac957}

\bibitem[{Rein \& Liu(2012)}]{Rein12}
Rein, H., \& Liu, S.-F. 2012, Astronomy \& Astrophysics, 537, A128

\bibitem[{{Ricker} {et~al.}(2015){Ricker}, {Winn}, {Vanderspek}, {Latham},
  {Bakos}, {Bean}, {Berta-Thompson}, {Brown}, {Buchhave}, {Butler}, {Butler},
  {Chaplin}, {Charbonneau}, {Christensen-Dalsgaard}, {Clampin}, {Deming},
  {Doty}, {De Lee}, {Dressing}, {Dunham}, {Endl}, {Fressin}, {Ge}, {Henning},
  {Holman}, {Howard}, {Ida}, {Jenkins}, {Jernigan}, {Johnson}, {Kaltenegger},
  {Kawai}, {Kjeldsen}, {Laughlin}, {Levine}, {Lin}, {Lissauer}, {MacQueen},
  {Marcy}, {McCullough}, {Morton}, {Narita}, {Paegert}, {Palle}, {Pepe},
  {Pepper}, {Quirrenbach}, {Rinehart}, {Sasselov}, {Sato}, {Seager},
  {Sozzetti}, {Stassun}, {Sullivan}, {Szentgyorgyi}, {Torres}, {Udry}, \&
  {Villasenor}}]{Ricker14}
{Ricker}, G.~R., {Winn}, J.~N., {Vanderspek}, R., {et~al.} 2015, Journal of
  Astronomical Telescopes, Instruments, and Systems, 1, 014003,
  \dodoi{10.1117/1.JATIS.1.1.014003}

\bibitem[{{Schlesinger}(1916)}]{1916PAllO...3...23S}
{Schlesinger}, F. 1916, Publications of the Allegheny Observatory of the
  University of Pittsburgh, 3, 23

\bibitem[{{Schmitt} \& {Vanderburg}(2021)}]{2021arXiv210310285S}
{Schmitt}, A., \& {Vanderburg}, A. 2021, arXiv e-prints, arXiv:2103.10285.
\newblock \doarXiv{2103.10285}

\bibitem[{{Schmitt} {et~al.}(2019){Schmitt}, {Hartman}, \&
  {Kipping}}]{2019arXiv191008034S}
{Schmitt}, A.~R., {Hartman}, J.~D., \& {Kipping}, D.~M. 2019, arXiv e-prints,
  arXiv:1910.08034.
\newblock \doarXiv{1910.08034}

\bibitem[{{Skrutskie} {et~al.}(2006){Skrutskie}, {Cutri}, {Stiening},
  {Weinberg}, {Schneider}, {Carpenter}, {Beichman}, {Capps}, {Chester},
  {Elias}, {Huchra}, {Liebert}, {Lonsdale}, {Monet}, {Price}, {Seitzer},
  {Jarrett}, {Kirkpatrick}, {Gizis}, {Howard}, {Evans}, {Fowler}, {Fullmer},
  {Hurt}, {Light}, {Kopan}, {Marsh}, {McCallon}, {Tam}, {Van Dyk}, \&
  {Wheelock}}]{2MASS}
{Skrutskie}, M.~F., {Cutri}, R.~M., {Stiening}, R., {et~al.} 2006, \aj, 131,
  1163, \dodoi{10.1086/498708}

\bibitem[{{Stassun} {et~al.}(2018){Stassun}, {Oelkers}, {Pepper}, {Paegert},
  {De Lee}, {Torres}, {Latham}, {Charpinet}, {Dressing}, {Huber}, {Kane},
  {L{\'e}pine}, {Mann}, {Muirhead}, {Rojas-Ayala}, {Silvotti}, {Fleming},
  {Levine}, \& {Plavchan}}]{TIC}
{Stassun}, K.~G., {Oelkers}, R.~J., {Pepper}, J., {et~al.} 2018, \aj, 156, 102,
  \dodoi{10.3847/1538-3881/aad050}

\bibitem[{{Sullivan} {et~al.}(2015){Sullivan}, {Winn}, {Berta-Thompson},
  {Charbonneau}, {Deming}, {Dressing}, {Latham}, {Levine}, {McCullough},
  {Morton}, {Ricker}, {Vanderspek}, \& {Woods}}]{2015ApJ...809...77S}
{Sullivan}, P.~W., {Winn}, J.~N., {Berta-Thompson}, Z.~K., {et~al.} 2015, \apj,
  809, 77, \dodoi{10.1088/0004-637X/809/1/77}

\bibitem[{{Tobin} {et~al.}(2016){Tobin}, {Kratter}, {Persson}, {Looney},
  {Dunham}, {Segura-Cox}, {Li}, {Chandler}, {Sadavoy}, {Harris}, {Melis}, \&
  {P{\'e}rez}}]{2016Natur.538..483T}
{Tobin}, J.~J., {Kratter}, K.~M., {Persson}, M.~V., {et~al.} 2016, \nat, 538,
  483, \dodoi{10.1038/nature20094}

\bibitem[{{Tokovinin}(2017)}]{Tokovinin2017}
{Tokovinin}, A. 2017, \mnras, 468, 3461, \dodoi{10.1093/mnras/stx707}

\bibitem[{{Tokovinin} \& {Moe}(2020)}]{2020MNRAS.491.5158T}
{Tokovinin}, A., \& {Moe}, M. 2020, \mnras, 491, 5158,
  \dodoi{10.1093/mnras/stz3299}

\bibitem[{{Toonen} {et~al.}(2016){Toonen}, {Hamers}, \& {Portegies
  Zwart}}]{Toonen2016}
{Toonen}, S., {Hamers}, A., \& {Portegies Zwart}, S. 2016, Computational
  Astrophysics and Cosmology, 3, 6, \dodoi{10.1186/s40668-016-0019-0}

\bibitem[{{Virtanen} {et~al.}(2020){Virtanen}, {Gommers}, {Oliphant},
  {Haberland}, {Reddy}, {Cournapeau}, {Burovski}, {Peterson}, {Weckesser},
  {Bright}, {van der Walt}, {Brett}, {Wilson}, {Jarrod Millman}, {Mayorov},
  {Nelson}, {Jones}, {Kern}, {Larson}, {Carey}, {Polat}, {Feng}, {Moore},
  {VanderPlas}, {Laxalde}, {Perktold}, {Cimrman}, {Henriksen}, {Quintero},
  {Harris}, {Archibald}, {Ribeiro}, {Pedregosa}, {van Mulbregt}, \&
  {Contributors}}]{scipy}
{Virtanen}, P., {Gommers}, R., {Oliphant}, T.~E., {et~al.} 2020, Nature
  Methods, \dodoi{https://doi.org/10.1038/s41592-019-0686-2}

\bibitem[{{Zasche} {et~al.}(2022){Zasche}, {Henzl}, \&
  {Ma{\v{s}}ek}}]{2022A&A...664A..96Z}
{Zasche}, P., {Henzl}, Z., \& {Ma{\v{s}}ek}, M. 2022, \aap, 664, A96,
  \dodoi{10.1051/0004-6361/202243723}

\bibitem[{{Zasche} {et~al.}(2023){Zasche}, {Borkovits}, {Jayaraman},
  {Rappaport}, {Bro{\v{z}}}, {Vokrouhlick{\'y}}, {B{\'\i}r{\'o}},
  {Heged{\"u}s}, {Kiss}, {Uhla{\v{r}}}, {Schwengeler}, {P{\'a}l},
  {Ma{\v{s}}ek}, {Howell}, {Dallaporta}, {Munari}, {Gagliano}, {Jacobs},
  {Kristiansen}, {LaCourse}, {Omohundro}, {Terentev}, {Vanderburg}, {Henzl},
  {Powell}, \& {Kostov}}]{2023MNRAS.tmp..346Z}
{Zasche}, P., {Borkovits}, T., {Jayaraman}, R., {et~al.} 2023, \mnras,
  \dodoi{10.1093/mnras/stad328}

\end{thebibliography}
\bibliographystyle{aasjournal}



\end{document}